%% file: main.tex
\documentclass{article}

\usepackage[linesnumbered,ruled,vlined]{algorithm2e}
\usepackage{microtype}
\usepackage{graphicx}
\usepackage{subfigure}
\usepackage{booktabs} 

\usepackage{hyperref}
\usepackage{multirow}

\usepackage[accepted]{icml2024}

\usepackage{amsmath}
\usepackage{amssymb}
\usepackage{mathtools}
\usepackage{amsthm}
\usepackage{bbm}

\usepackage{amssymb}
\usepackage{pifont}
\newcommand{\xmark}{\ding{55}}%

\usepackage{wrapfig}

\usepackage[capitalize,noabbrev]{cleveref}

\SetKwProg{kwServer}{ServerUpdate}{:}{}
\SetKwProg{kwClient}{ClientUpdate}{:}{}
\SetKwProg{kwServer}{\textcolor{blue}{Server Update}}{}{}
\SetKwProg{kwClient}{\textcolor{blue}{Client Update}}{}{}
\SetKwInput{Output}{Output}
\SetKwInput{Input}{Input}
\SetKwInput{Return}{Return}

\usepackage[textsize=tiny]{todonotes}
\newcommand{\ours}{\texttt{FedType}\xspace}
\newcommand{\model}{\texttt{FedType}}
\usepackage{xspace}

\icmltitlerunning{Bridging Model Heterogeneity in Federated Learning via Uncertainty-based
Asymmetrical Reciprocity Learning}

\begin{document}

\twocolumn[
\icmltitle{Bridging Model Heterogeneity in Federated Learning via Uncertainty-based Asymmetrical Reciprocity Learning}



\icmlsetsymbol{equal}{*}

\begin{icmlauthorlist}
\icmlauthor{Jiaqi Wang}{yyy}
\icmlauthor{Chenxu Zhao}{sch}
\icmlauthor{Lingjuan Lyu}{comp}
\icmlauthor{Quanzeng You}{cmp}
\icmlauthor{Mengdi Huai}{sch}
\icmlauthor{Fenglong Ma}{yyy}

\end{icmlauthorlist}

\icmlaffiliation{yyy}{Pennsylvania State University}
\icmlaffiliation{comp}{Sony AI}
\icmlaffiliation{cmp}{ByteDance}
\icmlaffiliation{sch}{Iowa State University}

\icmlcorrespondingauthor{Fenglong Ma}{fenglong@psu.edu}

\icmlkeywords{Federated Learning, Model Heterogeneity}

\vskip 0.3in
]



\printAffiliationsAndNotice{} 

\begin{abstract}
This paper presents \ours, a simple yet pioneering framework designed to fill research gaps in heterogeneous model aggregation within federated learning (FL). \ours introduces small identical proxy models for clients, serving as agents for information exchange, ensuring model security, and achieving efficient communication simultaneously. To transfer knowledge between large private and small proxy models on clients, we propose a novel uncertainty-based asymmetrical reciprocity learning method, eliminating the need for any public data.
Comprehensive experiments conducted on benchmark datasets demonstrate the efficacy and generalization ability of \ours across diverse settings. Our approach redefines federated learning paradigms by bridging model heterogeneity, eliminating reliance on public data, prioritizing client privacy, and reducing communication costs.
\end{abstract}

\input{section/introduction}
\input{section/related_work}

\input{section/method_new}

\input{section/experiment}
\input{section/conclusion}
\label{submission}

\nocite{langley00}

\bibliography{main}
\bibliographystyle{icml2024}

\input{section/appendix}

You can have as much text here as you want. The main body must be at most $8$ pages long.
For the final version, one more page can be added.
If you want, you can use an appendix like this one.  

The $\mathtt{\backslash onecolumn}$ command above can be kept in place if you prefer a one-column appendix, or can be removed if you prefer a two-column appendix.  Apart from this possible change, the style (font size, spacing, margins, page numbering, etc.) should be kept the same as the main body.

\end{document}

%% file: section/introduction.tex
\vspace{-0.1in}
\section{Introduction}\label{sec:intro}


Federated Learning (FL) is designed to enable the collaborative training of a machine learning model without the need to share clients' data. Many prevalent FL models, including FedAvg~\cite{mcmahan2017communication} and FedProx~\cite{li2020federated}, mandate that clients employ an identical model structure and target for training a shared global model. However, clients may possess diverse model structures, introducing \textbf{model heterogeneity} within the FL framework. The goal of this challenging task is to learn personalized client models instead of a powerful global model.

Recently, many studies have emerged to tackle the challenge of model heterogeneity~\cite{huang2022learn,li2019fedmd,yi2023fedgh,lin2020ensemble, yu2022resource,wang2023towards}. These efforts can be categorized into two groups based on the approach employed for information exchange between clients and the server. The first category focuses on the transmission of additional side information, such as logits~\cite{huang2022learn}, class scores~\cite{li2019fedmd}, and label-wise representations~\cite{yi2023fedgh,tan2022fedproto}, which are derived from utilizing public data on individual clients. Conversely, the second category involves the direct upload of client models to the server for processes like distillation~\cite{lin2020ensemble, yu2022resource} or model reassembly~\cite{wang2023towards} with the help of public data. While they successfully achieve the goal of heterogeneous model aggregation, they encounter significant drawbacks:

\textbf{Diminishing returns in public data incorporation} -- As previously mentioned, prevailing approaches rely on extra public data, either at the client or server side, for conducting heterogeneous model aggregation. The latest study~\cite{wang2023towards} underscores that the choice of public data plays a pivotal role in influencing model performance. Furthermore, using extra data amplifies the learning cost for models. To mitigate these challenges, a pertinent research question arises: \emph{Can heterogeneous models be successfully aggregated without depending on any external public data?}

\textbf{Disclosure risks raised by exchanging sensitive information} -- While exchanging side information between clients and the server can alleviate communication costs, this straightforward approach raises concerns about the potential disclosure of sensitive client information~\cite{lyu2022privacy}. Furthermore, the complete upload of model structures and parameters poses substantial security risks~\cite{tolpegin2020data} and privacy concerns~\cite{bouacida2021vulnerabilities}, particularly for business corporations or entities operating in sensitive domains.
In light of these challenges, another pivotal research question surfaces: \emph{Can heterogeneous model aggregation be achieved while exchanging only non-sensitive information, mitigating the risks associated with data privacy and security?}

\begin{figure*}[t!]
\centering
\includegraphics[width=0.88\textwidth]{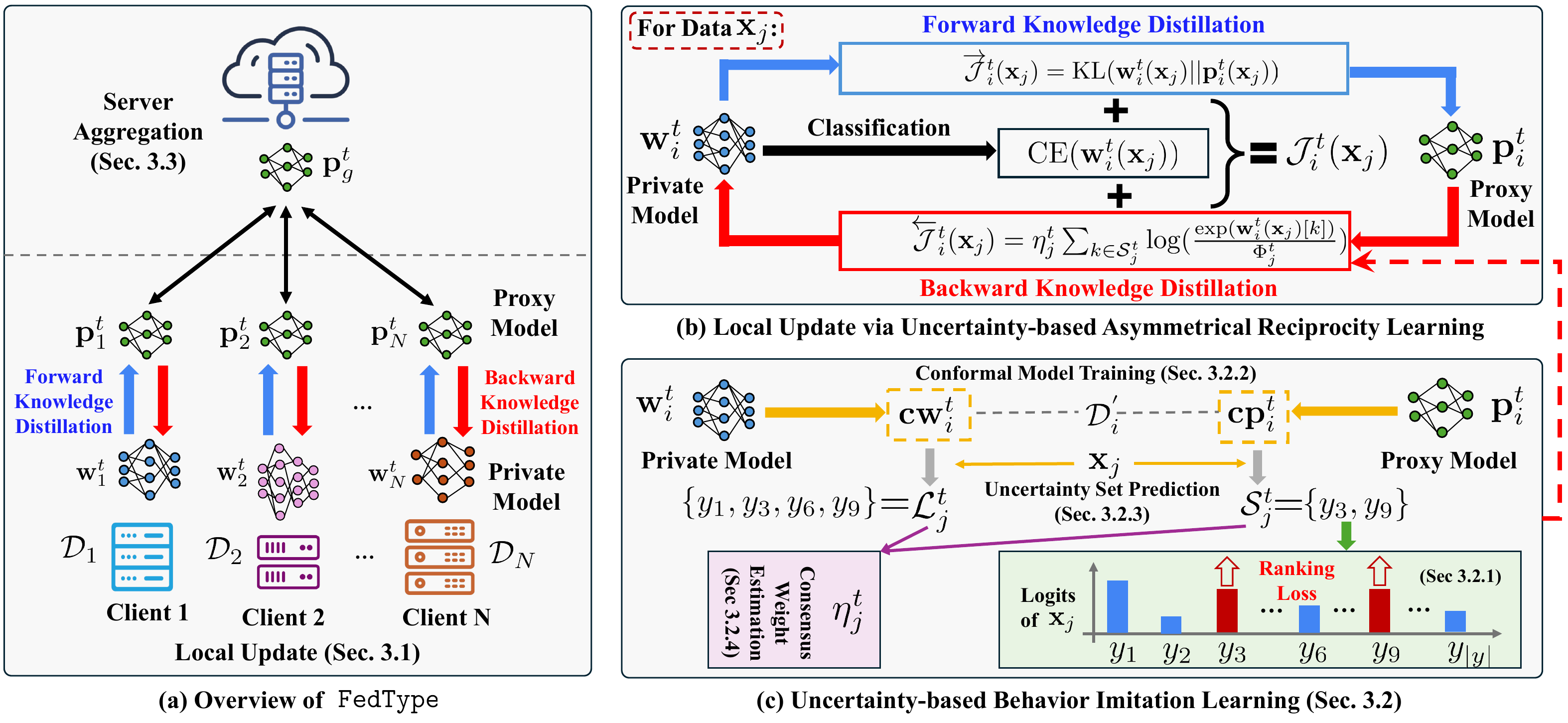}
    \vspace{-0.1in}
    \caption{Overview of the proposed \ours framework. (\textbf{a}) demonstrates the workflow of the proposed \ours to address the model-heterogeneous issue in FL, (\textbf{b}) is the local update demonstration for a data sample $\mathbf{x}_j$ using the proposed uncertainty-based asymmetrical reciprocity learning, and (\textbf{c}) is the illustration of backward knowledge distillation with the proposed uncertainty-based behavior imitation learning.}
    \label{fig:overview}
     \vspace{-0.15in}
\end{figure*}

\textbf{Necessity of efficient communication} -- Efficient communication stands out as a foundational challenge in federated learning. While numerous effective solutions have been devised for homogeneous federated learning~\cite{diao2020heterofl,yao2021fedhm, cho2023communication}, a notable research gap persists in heterogeneous federated learning.
While the exchange of side information effectively reduces communication costs, as discussed previously, these approaches introduce vulnerabilities to FL systems. Hence, an urgent requirement arises for an efficient, secure solution that operates without relying on public data in heterogeneous FL.

To resolve the issues mentioned previously, this paper introduces a novel, powerful, yet straightforward framework, \ours, to bridge model heterogeneity gaps in \underline{\textbf{fed}}erated learning without relying on any public data and ensuring efficient communication through uncertain\underline{\textbf{t}}y-based as\underline{\textbf{y}}mmetrical reci\underline{\textbf{p}}rocity l\underline{\textbf{e}}arning, as depicted in Figure~\ref{fig:overview}. \ours achieves heterogeneous model aggregation, communication efficiency, and system security simultaneously.


\ours involves introducing small identical proxy models for clients, serving as agents for information exchange between clients and the server to facilitate heterogeneous model aggregation. In the local training phase, each client employs a novel bidirectional knowledge distillation strategy, referred to as \textbf{asymmetrical reciprocity learning}, to simultaneously update its large private model and small proxy model. This strategy leverages the distinctive characteristics of these two models, elaborated in Section~\ref{sec:UARL}.
Additionally, an \emph{uncertainty-based behavior imitation learning} method is developed to enhance the guidance provided by the small proxy model in facilitating the learning of the large client model, as detailed in Section~\ref{sec:bkd_loss}.
Ultimately, each local client is mandated to upload only its proxy model to the server. This facilitates model aggregation using existing approaches employed in homogeneous FL, as outlined in Section~\ref{sec:server_update}.
We perform comprehensive experiments on multiple benchmark datasets, evaluating the \ours framework in both heterogeneous and homogeneous scenarios against state-of-the-art baselines. The experimental results consistently affirm the effectiveness of the \ours framework, underscoring its practicality and robust performance in real-world applications.

Remarkably, the \ours framework\footnote{The code is available at \url{https://github.com/JackqqWang/FedType}.} presents several notable advantages in comparison to existing approaches. Firstly, it eliminates the necessity for relying on any public data across clients and the server, effectively mitigating the adverse effects associated with the initialization of public data in model-heterogeneous federated learning. Secondly, \ours adopts an identical yet compact proxy model to facilitate information exchange between clients and the server, resulting in a substantial reduction in communication costs. Thirdly, by utilizing proxy models, the framework ensures the privacy and security of client models, offering protection against potential harmful attacks. Fourthly, the proposed framework exhibits versatility and can be applied to both heterogeneous and homogeneous federated learning settings. Lastly, this flexible framework is compatible with any existing federated learning model. 

%% file: section/related_work.tex
\vspace{-0.1in}

\section{Related Work}

\subsection{Heterogeneous Federated Learning}

Most existing federated learning studies~\cite{mcmahan2017communication,t2020personalized,zhang2022personalized, marfoq2022personalized,bao2023optimizing,dennis2021heterogeneity,DBLP:conf/cikm/Zhou0WH22} have concentrated on the homogeneous setting, requiring all clients to adopt an identical model structure.
Recent research has begun exploring the heterogeneous setting, allowing for varying model structures across clients. Within this domain, the approaches include sub-model training~\cite{alam2022fedrolex}, sparse model-adaption~\cite{chen2023efficient}, and hypernetworks~\cite{shamsian2021personalized}, which impose constraints on the relationship between the clients' models and the global model, thus still restricting the freedom of clients in using their preferred models. 

To facilitate more flexible cooperation among heterogeneous models, several studies have explored alternatives to averaging model parameters, such as aggregating extra information like logits~\cite{huang2022learn}, class scores~\cite{li2019fedmd}, and label-wise representations~\cite{yi2023fedgh,tan2022fedproto}. However, these methods potentially raise privacy concerns. In response, recent research has focused on exchanging model parameters through strategies like ensemble learning~\cite{lin2020ensemble}, mutual learning~\cite{yu2022resource,shen2023federated}, or model reassembly~\cite{wang2023towards}. 


These approaches exhibit two prevalent limitations: (1) Their dependence on public data. In real-world applications, obtaining access to public data may not always be feasible, and selecting suitable public data without preliminary knowledge of the client's data poses a complex challenge. (2) All the mentioned studies require sharing sensitive information, such as model structure, model parameters, or data-related insights from the private local models, with the server. This raises significant privacy concerns.

\vspace{-0.1in}
\subsection{Conformal Prediction}
Conformal prediction has seen significant popularity in recent years, particularly due to its capacity to generate prediction sets with guaranteed error rates under minimal assumptions. Initially introduced in ~\cite{vovk1999machine} and further elaborated in~\cite{shafer2008tutorial} and~\cite{balasubramanian2014conformal}, conformal prediction offers a distribution-free uncertainty quantification technique that has been effectively applied in various applications~\cite{angelopoulos2021gentle, bhatt2021uncertainty,fisch2021few,sankaranarayanan2022semantic}. 
Several recent studies focus on adapting conformal prediction within FL, specifically targeting the challenges associated with label shift~\cite{plassier2023conformal} and the quantification of uncertainty in distributed environments~\cite{lu2023federated}. To the best of our knowledge, there is limited work utilizing conformal prediction to enhance the performance of heterogeneous FL frameworks.




%% file: section/method_new.tex
\vspace{-0.1in}
\section{Methodology}

As illustrated in Figure~\ref{fig:overview}(a), \ours comprises two key components: local update and server update. During the $t$-th communication round, \ours initiates the training process for a proxy model $\mathbf{p}_i^t$ and a client private model $\mathbf{w}_i^t$ using the data $\mathcal{D}_i$ from the $i$-th client. It is essential to note that in model-heterogeneous federated learning, the structures of private models $\{\mathbf{w}_1^t, \cdots, \mathbf{w}_N^t\}$ differ from one another, where $N$ represents the number of clients. Conversely, the structures of proxy models $\{\mathbf{p}_1^t, \cdots, \mathbf{p}_N^t\}$ are uniform. Furthermore, it is worth mentioning that the parameter size of proxy model $\mathbf{p}_i^t$ is typically much smaller than that of private model $\mathbf{w}_i^t$. 

Only the learned proxy models $\{\mathbf{p}_1^t, \cdots, \mathbf{p}_N^t\}$\footnote{In the cross-device setting, we will randomly select $B \ll N$ clients at each communication round.} are transmitted to the server for model aggregation during the server update, leading to a significant reduction in communication costs.
The resulting aggregated model $\mathbf{p}_g^t$ is then disseminated to each client in the subsequent communication round as the initialized proxy model, denoted as $\mathbf{p}_i^{t+1} = \mathbf{p}_g^t$ ($\forall i \in [1, \cdots, N]$). These two updates are executed iteratively until \ours converges. The whole algorithm flow can be found in Appendix \ref{apd:algorithm}. Next, we provide the details of each update.
\vspace{-0.1in}
\subsection{Local Update via Uncertainty-based Asymmetrical Reciprocity Learning}\label{sec:UARL}
The inherent challenge posed by model heterogeneity renders the direct aggregation of uploaded client models unfeasible. Despite various proposed approaches for heterogeneous model aggregation~\cite{lin2020ensemble, yu2022resource, huang2022learn, li2019fedmd,yi2023fedgh}, as discussed in Section~\ref{sec:intro}, they still exhibit several limitations. In contrast to existing work, we present a simple yet novel uncertainty-based asymmetrical reciprocity learning (\texttt{UARL}) approach to tackle the challenges posed by model heterogeneity. 
As shown in Figure~\ref{fig:overview} (c), \texttt{UARL} is a bidirectional knowledge distillation (KD)-based model. 

The forward KD (FKD) follows the traditional ``teacher-student'' knowledge distillation approach. In FKD, a small proxy/student model $\mathbf{p}_i^t$ is distilled from the large, private client/teacher model $\mathbf{w}_i^t$ using the client data $\mathcal{D}_i$ for the $i$-th client during the $t$-th communication round. The forward loss can be formulated as follows:
\begin{equation}\label{eq:fkd}
    \overrightarrow{\mathcal{J}}_i^t = \sum_{j=1}^{|\mathcal{D}_i|}\text{KL}(\mathbf{w}_i^t(\mathbf{x}_{j})|| \mathbf{p}_i^t(\mathbf{x}_{j})),
\end{equation}
where $|\mathcal{D}_i|$ denotes the number of data stored in the $i$-th client, $\text{KL}(\cdot, \cdot)$ is the Kullback–Leibler divergence, and $\mathbf{x}_j \in \mathcal{D}_i$ represents the input data.

The backward KD (BKD) poses a challenge, as the capability of the small proxy model $\mathbf{p}_i^t$ is typically weaker than that of the large client model $\mathbf{w}_i^t$. Directly applying traditional knowledge distillation may lead to a degradation in the power of client models. To overcome this asymmetrical reciprocity issue, we introduce a novel uncertainty-based behavior imitation learning method to transfer diverse knowledge from the proxy model $\mathbf{p}_i^t$ to the client model $\mathbf{w}_i^t$. Further details can be found in Section~\ref{sec:bkd_loss}.

Let $\overleftarrow{\mathcal{J}}_i^t$ denote the backward KD loss. 
We then use the following loss function to train the $i$-th client at the $t$-th communication round:
\begin{equation}\label{eq:client_training}
    \mathcal{J}_i^t = \sum_{j=1}^{|\mathcal{D}_i|}\text{CE}(\mathbf{w}_i^t(\mathbf{x}_j), \mathbf{y}_j) + \overrightarrow{\mathcal{J}}_i^t + \overleftarrow{\mathcal{J}}_i^t,
\end{equation}
where $\mathbf{y}_j$ denotes the vectorized ground truth obtained from the real label set $\mathcal{Y}_j$ of $\mathbf{x}_j$. The algorithm flow can be found in Algorithm~\ref{alg:uarl}.

\begin{algorithm}[t]
\small
\SetAlgoLined
\DontPrintSemicolon

\Input{Client training data $\mathcal{D}_i$, private model $\mathbf{w}_i^{t-1}$, proxy model initialized by $\mathbf{p}_g^{t-1}$, validation data $\mathcal{D}_i^\prime$, local training epoch $R$, hyperparameters}

{Divide the shuffled $\mathcal{D}_i$ into $R$ parts $\{\mathcal{D}_i^1, \cdots, \mathcal{D}_i^R\}$;}

{Initialize epoch-level models: $\mathbf{w}_i^0 = \mathbf{w}_i^{t-1}$ and $\mathbf{p}_i^0 = \mathbf{p}_g^{t-1}$;}

\For{\textup{each epoch} $r = 1, \cdots, R$}{

    {Train the conformal prediction models $\mathbf{cw}_i^r$ and $\mathbf{cp}_i^r$ using the learned epoch-level models $\mathbf{w}_i^{r-1}$ and $\mathbf{p}_i^{r-1}$ with the validation data $\mathcal{D}_i^\prime$ according to Eq.~\eqref{eq:conformal_model};}

    {Initialize the loss $\mathcal{J}_i^r = 0$;}

    \For{\textup{each sample} $\mathbf{x}_j \in \mathcal{D}_i^r$}{
        {Calculate the classification loss via $\text{CE}(\mathbf{w}_i^{r-1}(\mathbf{x}_j), \mathbf{y}_j)$;}

        {Calculate the FKD loss $\overrightarrow{\mathcal{J}}_i^r(\mathbf{x}_j)$ via Eq.~\eqref{eq:fkd};}

        {Calcuate the performance change $\Delta^r$ on the proxy model on $\mathcal{D}_i^\prime$;}
        
        {Obtain the prediction set $\mathcal{S}_j^r$ via Eq.~\eqref{eq:cp} using $\mathbf{cp}_i^r$;}
        
        {Obtain the prediction set $\mathcal{L}_j^r$ via Eq.~\eqref{eq:cp} using $\mathbf{cw}_i^r$;}

        {Calculate $\eta_j^r$ according to Eq.~\eqref{eq:eta};}

        {Calculate the BKD loss $\overleftarrow{\mathcal{J}}_i^r(\mathbf{x}_j)$;}

        {$\mathcal{J}_i^r$ += $\text{CE}(\mathbf{w}_i^{r-1}(\mathbf{x}_j), \mathbf{y}_j) + \overrightarrow{\mathcal{J}}_i^r(\mathbf{x}_j) + \overleftarrow{\mathcal{J}}_i^r(\mathbf{x}_j)$;}
    }

    {Update the models $\mathbf{w}_i^{r}$ and $\mathbf{p}_i^{r}$ by optimizing $\mathcal{J}_i^r$;}
}

\Return{Trained models $\mathbf{w}_i^t$ and $\mathbf{p}_i^t$.}

\caption{Epoch-Level Algorithm Flow of \texttt{UARL}.}
\label{alg:uarl}
\end{algorithm}
\vspace{-0.1in}


\subsection{Uncertainty-based Behavior Imitation Learning}\label{sec:bkd_loss}
The initialization of the proxy model $\mathbf{p}_i^t$ is carried out using the aggregated global model $\mathbf{p}_g^{t-1}$ (refer to Section~\ref{sec:server_update}). This global model encapsulates diverse knowledge from other clients, making it crucial to transfer this knowledge to the private client model. However, due to the inherently weaker capability of the proxy model $\mathbf{p}_i^t$, direct application of the traditional ``teacher-student'' learning paradigm for knowledge transfer to the large model $\mathbf{w}_i^t$ is impractical. Such a direct approach may introduce additional noise to the large model, potentially impeding the overall training efficiency of the entire framework.

\subsubsection{Backward Knowledge Distillation Loss}
To tackle this challenge, we introduce an uncertainty-based behavior imitation learning approach. This method exclusively relies on the use of \emph{partial logits} generated by the proxy model $\mathbf{p}_i^t$ with a high level of confidence or certainty. The intention is to employ these confident predictions to guide the learning process of the large model $\mathbf{w}_i^t$.
In essence, when dealing with a specific data sample $\mathbf{x}_j \in \mathcal{D}_i$, if the proxy model $\mathbf{p}_i^t$ exhibits high confidence in predicting certain classes, transferring this behavioral information—rather than the complete logits—can still be beneficial for the training of the large model.

Let $\mathcal{S}_j^t$ represent the set of class labels predicted with high confidence by $\mathbf{p}_i^t$. In our proposed behavior imitation learning, the objective is to strengthen the probability/logit associated with the labels in $\mathcal{S}_j^t$ as predicted by the large model $\mathbf{w}_i^t$. Importantly, we refrain from imposing a strict requirement for the large model $\mathbf{w}_i^t$ to prioritize ranking the labels in $\mathcal{S}_j^t$ at the top positions. To address this, we introduce a novel ranking-based behavior imitation learning loss, outlined as follows:
\begin{equation}\label{eq:certainty_KD}
\begin{split}
    &\overleftarrow{\mathcal{J}}_i^t = \sum_{j=1}^{|\mathcal{D}_i|} \eta_j^t \sum_{k \in \mathcal{S}_j^t} \log(\frac{\exp(\mathbf{w}_i^t(\mathbf{x}_j)[k])}{\Phi_j^t}),\\
    \Phi_j^t = &\sum_{s \in \mathcal{S}_j^t} \exp(\mathbf{w}_i^t(\mathbf{x}_j)[s]) + \sum_{v \in \mathcal{V}_j^t} \exp(\mathbf{w}_i^t(\mathbf{x}_j)[v]),
\end{split}
\end{equation}
where $\eta_j^t$ is the estimated consensus weight to determine the amount of the transferred knowledge from the proxy model $\mathbf{p}_i^t$ to the large model $\mathbf{w}_i^t$.
$\mathcal{V}_j^t = \mathcal{Y} - \mathcal{S}_j^t$ denotes the class labels associated with low confidence, and $\mathcal{Y}$ represents the complete set of class labels in $\mathcal{D}_i$. 
In Eq.~\eqref{eq:certainty_KD}, $\eta_j^t$ and $\mathcal{S}_j^t$ are unknown variables. 
To proceed, we introduce a novel approach to quantify an uncertainty set $\mathcal{S}_j^t$ for each data $\mathbf{x}_j$ through dynamic conformal prediction, and the uncertainty sets will be further used to estimate $\eta_j^t$.

\subsubsection{Conformal Model Training}

Conformal prediction~\cite{angelopoulos2021gentle}
stands out as a reliable and interpretable approach for quantifying uncertainty, providing prediction sets accompanied by a designated level of confidence or probability. In mathematical terms, conformal prediction involves an uncertainty set function $f(\mathbf{p}_j^t, \mathbf{x}_j)$ that maps $\mathbf{x}_j$ to a subset of $\mathcal{Y}$ (i.e., $f(\mathbf{p}_j^t, \mathbf{x}_j) = \mathcal{S}_j^t {\subseteq} \mathcal{Y}$), satisfying the condition:
\begin{equation}
    P(\mathcal{Y}_j \in \mathcal{S}_j^t)) \geq 1-\theta,
\end{equation}
where $\theta$ represents a predefined confidence level. 

To estimate the prediction set $\mathcal{S}_j^{t}$, we need to train a conformal model $\mathbf{cp}_j^t$ first using the validation dataset $\mathcal{D}_i^\prime$ where $\mathcal{D}_i \cap \mathcal{D}_i^\prime = \emptyset$ as follows:
\begin{equation}\label{eq:conformal_model}
    \mathbf{cp}_i^t = \text{Cmodel}(\mathbf{p}_i^t, \mathcal{D}_i^\prime),
\end{equation}
where $\text{Cmodel}()$ is constructed using the split conformal prediction framework, detailed in Appendix~\ref{apd:conformal}.

Existing conformal prediction approaches are principally developed to quantify uncertainty in static, well-trained models and are not tailored to address dynamic scenarios.
However, the \ours framework undergoes iterative training, resulting in dynamic changes to the models $\mathbf{p}_i^t$ and $\mathbf{w}_i^t$ at each communication round and even each epoch $r$.\footnote{In our implementation, the parameters of two models change at each epoch $r$ during each communication round $t$ in the training stage. Here, we omit the notation of epoch $r$ in the rest of this section for simplicity and readability, which can be treated as $r=1$ in each communication round $t$.} Consequently, the uncertainty set $\mathcal{S}_j^{t}$ is predicted dynamically.
Next, we will use the trained conformal model $\mathbf{cp}_i^t$ to generate the prediction set $\mathcal{S}_j^{t}$ for each training data $\mathbf{x}_j \in \mathcal{D}_i$.

\subsubsection{Uncertainty Set Prediction}\label{sec:usp}


A straightforward approach would involve directly applying existing conformal prediction methods, such as regularized adaptive prediction sets (RAPS)~\cite{angelopoulos2020uncertainty}, to generate the uncertainty set $\mathcal{S}_j^{t}$ for $\mathbf{x}_j$ at each communication round $t$. However, as mentioned earlier, the capability of the proxy model $\mathbf{p}_i^t$ is weak, especially at the initial stages of training. Utilizing prediction sets from such unreliable models may introduce adverse effects when training Eq.~\eqref{eq:certainty_KD}.


To address this challenge, we introduce a dynamic adjustment mechanism for the size of the prediction set $\mathcal{S}_j^{t}$ based on the observed changes in model performance on the training data $\mathcal{D}_i$. In essence, a decrease in a model's performance may be indicative of low-quality prediction sets. Therefore, to enhance the informativeness of the prediction sets, it is necessary to reduce their size by refining the conformal prediction algorithm.
Based on this intuition, we propose a dynamic conformal prediction for federated learning training based on RAPS using the following uncertainty prediction set generation:
\begin{equation}\label{eq:cp}
    \begin{split}
        \mathcal{S}_j^t &:= 
        \{y : \pi_j(y) \cdot u + \rho_j(y) \\
        &+ g(\Delta^t, \lambda) \cdot (o_j(y)-\kappa_{reg})^{+} \leq \tau\}.
    \end{split}
\end{equation}
$\pi_j(y)$ represents the probability assigned to the label $y$ as predicted by the conformal model $\mathbf{cp}_i^{t}$ for the $j$-th data point $\mathbf{x}_j$. The parameter $u$ is a predefined randomized factor determining the value jump for each new label $y$.
Furthermore, $\rho_j(y) = \sum_{y^\prime=1}^{|\mathcal{Y}|} \pi_j(y^\prime) \mathbbm{1}_{\{\pi_j(y^\prime) > \pi_j(y)\}}$ denotes the total probability mass associated with the set of labels that are more likely than the label $y$.

Besides, $g(\Delta^t, \lambda)$ is a piecewise calibration function\footnote{Here, we define a simple linear function to characterize the dynamic adjustment of prediction sets. We can also use other monotonically decreasing functions as alternative solutions.}, which is defined as follows:
\begin{equation}
    g(\Delta^t, \lambda) = 
    \begin{cases}
        \lambda \cdot \Delta^t - \Delta^t + \lambda, &\text{if $\Delta^t < 0$,}\\
        \lambda, &\text{otherwise}.
    \end{cases}
\end{equation}
$\Delta^t = \mathcal{A}(\mathbf{p}_i^{t}, \mathcal{D}_i^\prime) - \mathcal{A}(\mathbf{p}_i^{t-1}, \mathcal{D}_i^\prime)$, where $\mathcal{A}(\mathbf{p}_i^{t}, \mathcal{D}_i^\prime)$ denotes the validation accuracy on $\mathcal{D}_i^\prime$ using the trained model $\mathbf{p}_i^{t}$. 
In the event of a performance drop, i.e., $\Delta^t < 0$, we actively reduce the size of the prediction set by adjusting the value of $g(\Delta^t, \lambda)$. On the other hand, if $\Delta^t \geq 0$, we maintain $g(\Delta^t, \lambda) = \lambda$, where $\lambda \geq 0$ serves as a predefined regularization hyperparameter.

In addition, $(z)^+$ denotes the positive part of $z$, and $\kappa_{reg}$ is another regularization hyperparameter. The variable $o_j(y)=|y^\prime \in \mathcal{Y}: \pi_j(y^\prime) \geq \pi_j(y)|$ represents the ranking of $y$ among the labels in $\mathcal{Y}$ based on the prediction probability $\pi_j$. Finally, $\tau$ denotes the cumulative sum of the sorted, penalized classifier scores. The details of the designed dynamic conformal prediction can be found in Appendix~\ref{apd:conformal}.

\subsubsection{Consensus Weight Estimation}\label{sec:weight}
In Eq.~\eqref{eq:certainty_KD}, $\eta_j^t$ is a key factor to control the behavior of knowledge transfer from the proxy model $\mathbf{p}_i^t$ to the large model $\mathbf{w}_i^t$. Intuitively, if the proxy model $\mathbf{p}_i^t$ is significantly confident on certain predicted labels on data $\mathbf{x}_j$ i.e., the small size of $\mathcal{S}_j^t$, then it is necessary to transfer such high-quality knowledge to the large model $\mathbf{w}_i^t$ as much as possible, as the guidance of model training. On the other hand, we need to reduce the amount of transferred knowledge with low quality, i.e., the uncertainty set $\mathcal{S}_j^t$ with large size. 

The quality of knowledge is a relative variable, which can be determined by the size of uncertainty sets predicted by both models $\mathbf{p}_i^t$ and $\mathbf{w}_i^t$ on $\mathbf{x}_j$ using Eq.~\eqref{eq:cp} with the conformal models $\mathbf{cp}_i^t$ and $\mathbf{cw}_i^t$ learned by Eq.~\eqref{eq:conformal_model}, denoted as $\mathcal{S}_j^t$ and $\mathcal{L}_j^t$, respectively. Mathematically, we define the consensus weight $\eta_j^t$ as follows:
\begin{equation}\label{eq:eta}
    \eta_j^t = 
    \begin{cases}
        |\mathcal{S}_j^t \cap \mathcal{L}_j^t|/|\mathcal{S}_j^t \cup \mathcal{L}_j^t|, & \text{if $|\mathcal{S}_j^t| \geq |\mathcal{L}_j^t|$},\\
        |\mathcal{S}_j^t \cap \mathcal{L}_j^t|/|\mathcal{S}_j^t|, & \text{if $|\mathcal{S}_j^t| <|\mathcal{L}_j^t|$},
    \end{cases}
\end{equation}

where $|\cdot|$ denotes the size of the set. We can observe that such a design encourages the proxy model to transfer confident knowledge to the large model.
\vspace{-0.1in}
\subsection{Server Update}\label{sec:server_update}
After training each client using Eq.~\eqref{eq:client_training}, we will obtain both proxy and private models. The proxy models $\{\mathbf{p}_1^t, \cdots, \mathbf{p}_N^t\}$ will be uploaded to the server to conduct aggregation, resulting in a share proxy global model $\mathbf{p}_g^t$ using any existing data-free aggregation approaches, such as FedAvg~\cite{mcmahan2017communication} and FedProx~\cite{li2020federated}. The global model $\mathbf{p}_g^t$ will be distributed to each client at the next communication round as the initialization of the proxy client model, i.e., $\mathbf{p}_i^{t+1} = \mathbf{p}_g^t$. The proposed \ours framework will be iteratively executed for the client update and server update until convergence.

%% file: section/experiment.tex
\begin{table*}[!t]
\centering
\vspace{-0.1in}
\caption{Performance (\%) comparison under the heterogeneous cross-device settings.}
 \resizebox{0.9\textwidth}{!}
{
\begin{tabular}{c|l|ccc|ccc|ccc} 
\toprule 

\multirow{2}{*}{\textbf{Agg. Method}} 
& \multirow{2}{*}{\textbf{Model}} & \multicolumn{3}{c|}{\textbf{FMNIST}}&
\multicolumn{3}{c|}{\textbf{CIFAR-10}} & \multicolumn{3}{c}{\textbf{CIFAR-100}}\\\cline{3-11}
& & $\alpha = 1$& $\alpha = 0.5$ & $\alpha = 0.1$  & $\alpha = 1$ & $\alpha = 0.5$ & $\alpha = 0.1$ & $\alpha = 1$ &$\alpha = 0.5$ & $\alpha = 0.1$\\
\midrule
\xmark&FedProto&85.05&87.66&89.04&76.59&78.17&82.96&58.03&66.31&68.60\\
\xmark&FML&86.54&90.71&92.63&80.80&85.24&88.58&58.77&66.90&68.74\\

\hline\hline

\multirow{3}{*}{FedAvg}
&\model$_{\text{global}}$ &84.11&83.93&81.32&66.40&63.39&58.17&38.36&38.17&35.45\\
&\model$_{\text{proxy}}$ &86.09&89.45&93.16&80.65&82.57&85.04&56.24&61.06&62.31\\
&\model$_{\text{private}}$ &\textbf{87.26}&\textbf{91.22}&\textbf{94.77}&\textbf{82.56}&\textbf{86.83}&\textbf{91.90}&\textbf{57.33}&\textbf{65.69}&\textbf{68.14}\\\hline
\multirow{3}{*}{FedProx}
&\model$_{\text{global}}$ &86.96&86.44&84.29&68.26&65.86&63.75&41.88&39.31&36.53 \\
&\model$_{\text{proxy}}$  &87.03&91.50&92.64&82.19&82.48&87.80&58.56&61.22&62.64\\
&\model$_{\text{private}}$&\textbf{87.65}&\textbf{93.84}&\textbf{94.98}&\textbf{83.69}&\textbf{86.92}&\textbf{92.03}&\textbf{59.18}	&\textbf{65.45}&	\textbf{68.37}
 \\\hline

\multirow{3}{*}{pFedMe}
&\model$_{\text{global}}$ &87.82&87.13&85.86&68.71&65.22&64.95&41.55&40.92&38.60 \\
&\model$_{\text{proxy}}$  &88.63&92.05&93.38&82.64&83.00&88.14&59.04&62.68&64.89\\
&\model$_{\text{private}}$ 	&\textbf{88.96}&\textbf{92.36}&	\textbf{94.86}&	\textbf{83.47}	&\textbf{87.24}&	\textbf{92.16}	&\textbf{59.78}&\textbf{67.07}	&\textbf{69.51}
\\\hline
\multirow{3}{*}{pFedBayes}
&\model$_{\text{global}}$ &88.20&87.85&86.04&68.41&66.87&63.32&43.73&41.24&38.72\\
&\model$_{\text{proxy}}$ &89.69&92.11&93.29&83.33&84.49&89.10&59.47&62.96&63.51\\
&\model$_{\text{private}}$ &\textbf{90.26}&\textbf{93.17}&\textbf{95.88}&\textbf{84.09}&\textbf{88.67}&\textbf{92.38}&\textbf{59.62}&\textbf{67.35}&	\textbf{69.60}
\\

\bottomrule 
\end{tabular}
}
\label{tab:hete}
\vspace{-0.1in}
\end{table*}
\vspace{-0.1in}
\section{Experiments}
\subsection{Experimental Setups}
\noindent\textbf{Data preparation.}
We assess the effectiveness of the proposed \ours approach through image classification tasks conducted in the cross-device scenario on FMNIST, CIFAR-10, and CIFAR-100 datasets, and cross-silo scenario on Fed-ISIC19 dataset~\cite{ogier2022flamby}. For the cross-device experiments, we follow existing work~\cite{yurochkin2019bayesian,hsu2019measuring} to set heterogeneity degrees by adjusting the Dirichlet distribution's concentration parameter $\alpha$. We set $\alpha = 1, 0.5, 0.1$, respectively in our experiments. The details of the data partition can be found in Appendix~\ref{apd:datasets}.


\noindent\textbf{Baselines.} 
In our scenario, clients employ distinct network structures, and we abstain from using public data. FML~\cite{shen2023federated} is the only baseline with the same setting as ours, which uses bidirectional knowledge distillation to learn the model. Besides, we employ 
FedProto~\cite{tan2022fedproto} as another baseline, which aggregates class prototypes instead of model parameters.

The outcomes of \ours consist of three models: a shared global model (\model$_\text{global} = \mathbf{p}_g^T$), a set of proxy client models (\model$_\text{proxy} = \{\mathbf{p}_1^T, \cdots, \mathbf{p}_N^T\}$), and a set of private client models (\model$_\text{private} = \{\mathbf{w}_1^T, \cdots, \mathbf{w}_N^T\}$), where $T$ represents the total number of communication rounds. Subsequently, we compare the average client accuracy across these three types of models.
Moreover, the inherent flexibility of model aggregation in \ours enables us to assess various representative aggregation approaches, including FedAvg~\cite{mcmahan2017communication}, FedProx~\cite{li2020federated}, pFedMe~\cite{t2020personalized}, and pFedBayes~\cite{zhang2022personalized}. More details of the baselines can be found in Appendix~\ref{apd:baseline}.



\noindent \textbf{Impletation details.} 
To replicate the model-heterogeneous scenario, we assemble a model pool comprising private client models, encompassing ResNet-18, ResNet-34, ResNet-50, RestNet-101, ResNet-152, VGG-11, VGG-13, VGG-16, and VGG-19, which will be randomly assigned to a client.
In our primary experiments, we designate ResNet-18 as the small proxy model, as presented in Table~\ref{tab:hete}. Additionally, we delve into alternative options for proxy model selection, detailed in Table~\ref{tab:proxy_model_study}. Our reported metric is the \textbf{average accuracy} on \emph{100 clients for the cross-device setting and 6 clients for the cross-silo setting}. More details about the model and hyperparameters can be found in Appendix~\ref{apd: model} and ~\ref{apd:implementation}, respectively.

\vspace{-0.1in}
\subsection{Results of the Heterogeneous Model Setting}
\subsubsection{Cross-device Evaluation}
Table~\ref{tab:hete} presents the average client accuracy across three image datasets, employing different aggregation methods under varying label heterogeneity distributions ($\alpha$'s). Observing the results, it is evident that the proposed \model$_\text{private}$ consistently outperforms both the shared global model \model$_\text{global}$ and the learned proxy model \model$_\text{proxy}$ across diverse aggregation methods. 

Particularly noteworthy is the global model's lower performance, especially in challenging tasks. This aligns with our design, where the shared global model primarily serves as an agent for information exchange in personalized federated learning. Despite sharing the same network structure as the global model, the proxy models, through uncertainty-based asymmetrical reciprocity learning, acquire valuable knowledge, contributing to their enhanced performance.
Furthermore, comparing different aggregation methods reveals that personalized approaches (pFedMe and pFedBayes) exhibit superior performance compared to general methods (FedAvg and FedProx), aligning with our expectations.

Notably, an increase in the value of $\alpha$ corresponds to an overall performance improvement for \model$_\text{proxy}$ and \model$_\text{private}$. This observation aligns with our data partition method, where training and testing data follow the same distribution. A larger $\alpha$ increases label categories for each client, rendering the classification task more challenging. However, the performance of \model$_\text{global}$ contradicts this trend. This is attributed to the shared global model serving as an average representation of all proxy client models, performing better in scenarios where data follows an independent and identical distribution, such as a large $\alpha$.


\subsubsection{Cross-silo Evaluation}
\begin{figure}[!t]
    \centering
    \includegraphics[width=0.4\textwidth]{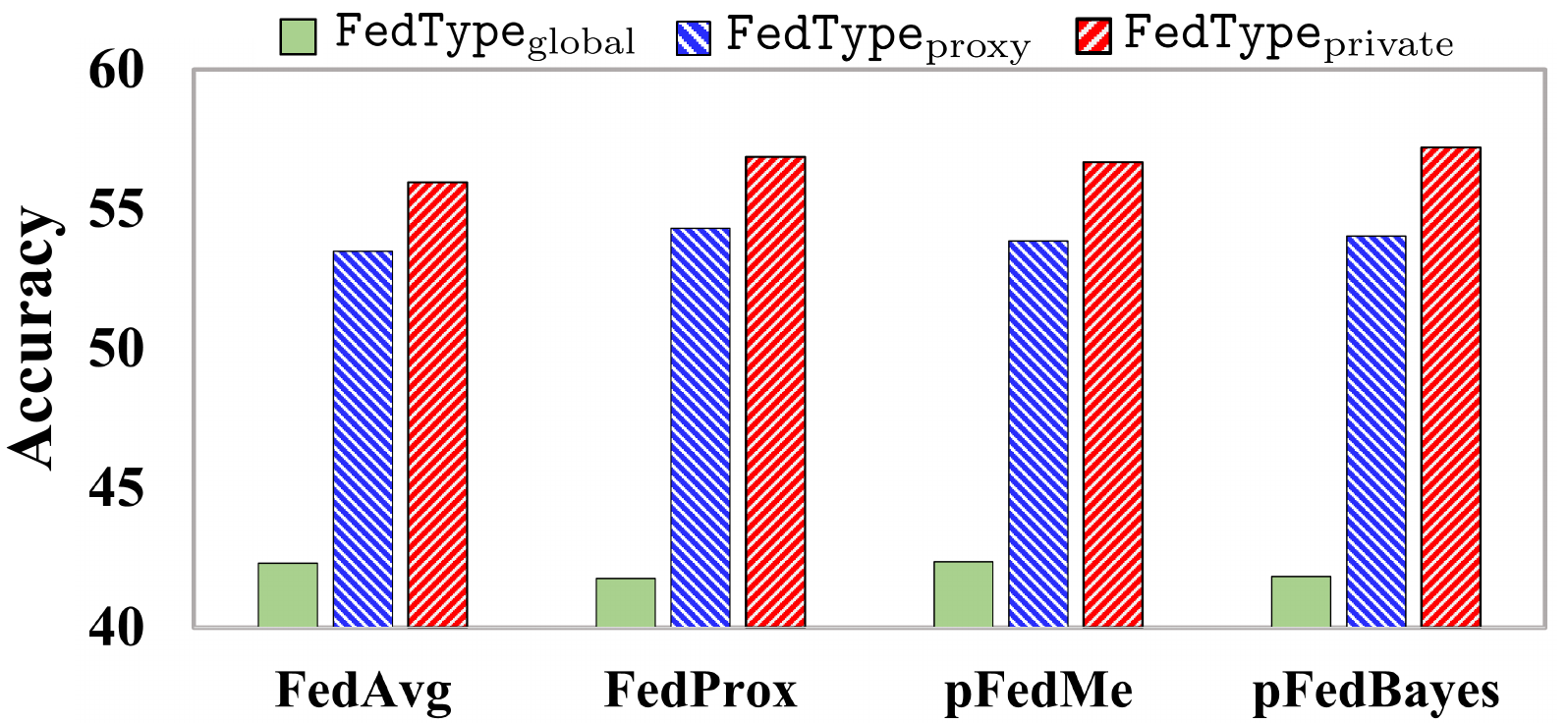}
    \vspace{-0.1in}
    \caption{Results (\%) of the cross-silo evaluation.}
    \label{fig:cross-silo}
    \vspace{-0.2in}
\end{figure}

An additional experiment is conducted to assess the effectiveness of the proposed \ours under the cross-silo setting, utilizing the Fed-ISIC19 dataset. The results are illustrated in Figure~\ref{fig:cross-silo}, where $x$-axis denotes the aggregation method, and $y$-axis is the average client accuracy. Similar observations to those in Table~\ref{tab:hete} emerge, where the private models outperform the proxy models, which, in turn, surpass the global models. This consistent trend validates the efficacy of the proposed uncertainty-based asymmetrical reciprocity learning in mutually enhancing the capabilities of proxy and private models.

\subsubsection{Ablation Study}
We compare our model with the following variants of \ours (i.e., \model$_\text{private}$) in this ablation study: 
(1) \model$_{\text{sym}}$: We replace the backward loss
$\overleftarrow{\mathcal{J}}_i^t$ with traditional knowledge distillation loss in Eq.~\eqref{eq:client_training}, which can be treated as symmetrical reciprocity learning. 
(2) \model$_{\text{Top}K}$: We choose top $K$ labels based on the ranking of the logits to construct the label set rather than using the proposed conformal prediction set estimation in the Section~\ref{sec:usp}. Here, we set $K = 3$. (3) \model$_{\eta=1}$: We simply set the consensus weight $\eta_j^t = 1$ in Eq.~\eqref{eq:eta}.
(4) \model$_{g=0.5}$: We simply set the function $g(\Delta^t, \lambda) = \lambda = 0.5$ in Eq.~\eqref{eq:cp}.
The ablation study results are shown in Table~\ref{tab:abl}. The cross-device scenario is validated on the CIFAR-10 dataset by setting $\alpha=0.5$ and the number of clients as 100. The aggregation method on the server is FegAvg for both cross-device and cross-silo scenarios.

Observing Table~\ref{tab:abl} reveals that each component effectively enhances performance but to varying degrees.
Firstly, \model$_{\text{sym}}$ exhibits the least favorable performance among all baselines, underscoring the imperative need for considering asymmetrical knowledge distillation.
Secondly, the outcomes of \model$_{\text{Top}K}$ suggest the effectiveness of asymmetrical reciprocity learning. However, opting for an arbitrarily fixed number of top ranks proves suboptimal. Therefore, it becomes essential to dynamically determine the ranks for each sample.
Thirdly, the results of \model$_{\eta=1}$ and \model$_{g=0.5}$ demonstrate that employing uncertainty set prediction contributes to a performance increase compared with \model$_{\text{Top}K}$.
In conclusion, considering asymmetrical reciprocity learning, dynamically adjusting the prediction set based on the model training performance, and estimating the consensus weights of samples prove to be valuable strategies for enhancing overall performance.


    



\begin{table}[t]
\label{tb:ablation}
\centering
\vspace{-0.1in}
\caption{Ablation study performance ($\%$) comparison.}

\resizebox{0.7\columnwidth}{!}{
\begin{tabular}{l|c|c} 
\toprule 

\textbf{Dataset}& \multicolumn{1}{c|}{\textbf{CIFAR-10}}& \multicolumn{1}{c}{\textbf{Fed-ISIC19}}\\
\midrule
\model$_{\text{sym}}$  &82.24&49.89\\
\model$_{\text{Top}K}$  & 84.69&51.26\\
\model$_{\eta=1}$  &85.12&52.40\\
\model$_{g=0.5}$  &86.24&54.17\\
\ours & \textbf{86.83}&\textbf{55.95}\\ 
\bottomrule 
\end{tabular}
}
\label{tab:abl}
\vspace{-0.2in}
\end{table}

\begin{table*}[!t]
\centering
\caption{Proxy model study. The approximate model sizes are shown using the model parameters (in millions).}
\resizebox{0.9\textwidth}{!}{
\begin{tabular}{c|c|ccc|ccc|ccc|c} 
\toprule 
Proxy& Parameter& \multicolumn{3}{c|}{\textbf{FMNIST}}&
\multicolumn{3}{c|}{\textbf{CIFAR-10}} & \multicolumn{3}{c|}{\textbf{CIFAR-100}} & \textbf{Fed-ISIC19}\\ \cline{3-12}
Model & Size   & $\alpha = 1$ & $\alpha = 0.5$ & $\alpha = 0.1$  & $\alpha = 1$& $\alpha = 0.5$ & $\alpha = 0.1$  & $\alpha = 1$& $\alpha = 0.5$ & $\alpha = 0.1$ & \xmark\\
\midrule


ShuffleNet-V2& 2.27M&87.11&88.86&91.58&81.25&81.79&84.48&50.44&57.17&60.35&47.48\\ 

MobileNet-V1&3.21M&87.59&88.93&91.13&81.04&82.95&85.06&51.31&59.30&62.76&50.32\\  

EfficientNet-B0& 5.29M&\textbf{88.43}&89.55&92.14&82.13&83.76&87.02&54.12&61.89&63.61&53.13\\ 


ResNet-18& 11.17M&87.26&\textbf{91.22}&\textbf{94.77}&\textbf{82.56}&\textbf{86.83}&\textbf{91.90}&\textbf{57.33}&\textbf{65.69}&\textbf{68.14} &\textbf{55.95}\\
\bottomrule 
\end{tabular}
}
\label{tab:proxy_model_study}
\end{table*}

\subsubsection{Dynamic Conformal Prediction}\label{sec:dcp}
In Eq.~\eqref{eq:cp}, we introduce a calibration function $g(\Delta^t, \lambda)$ to regulate the size of prediction sets based on the model's performance change. 
In this experiment, our objective is to examine the impact of selecting the function  $g(\Delta^t, \lambda)$ when $\Delta^t \in [-1, 0]$. 
Let $g_1 = g(\Delta^t, \lambda) = \lambda \cdot \Delta^t - \Delta^t + \lambda$ and $g_2=g(\Delta^t, \lambda) =\lambda$ ($\forall \Delta^t \in [-1, 1]$).
We also consider two alternative quadratic functions: $g_3=g(\Delta^t, \lambda) = \lambda \cdot {\Delta^t}^{2} + \lambda$ and $g_4=g(\Delta^t,\lambda) = -\lambda \cdot {\Delta^t}^{2} - {\Delta^t}+\lambda $. When $\Delta^t \in (0, 1]$, $g(\Delta^t, \lambda) =\lambda $ for all functions with $\lambda=0.5$.

\begin{figure}[t]
    \centering
    \includegraphics[width=0.45\textwidth]{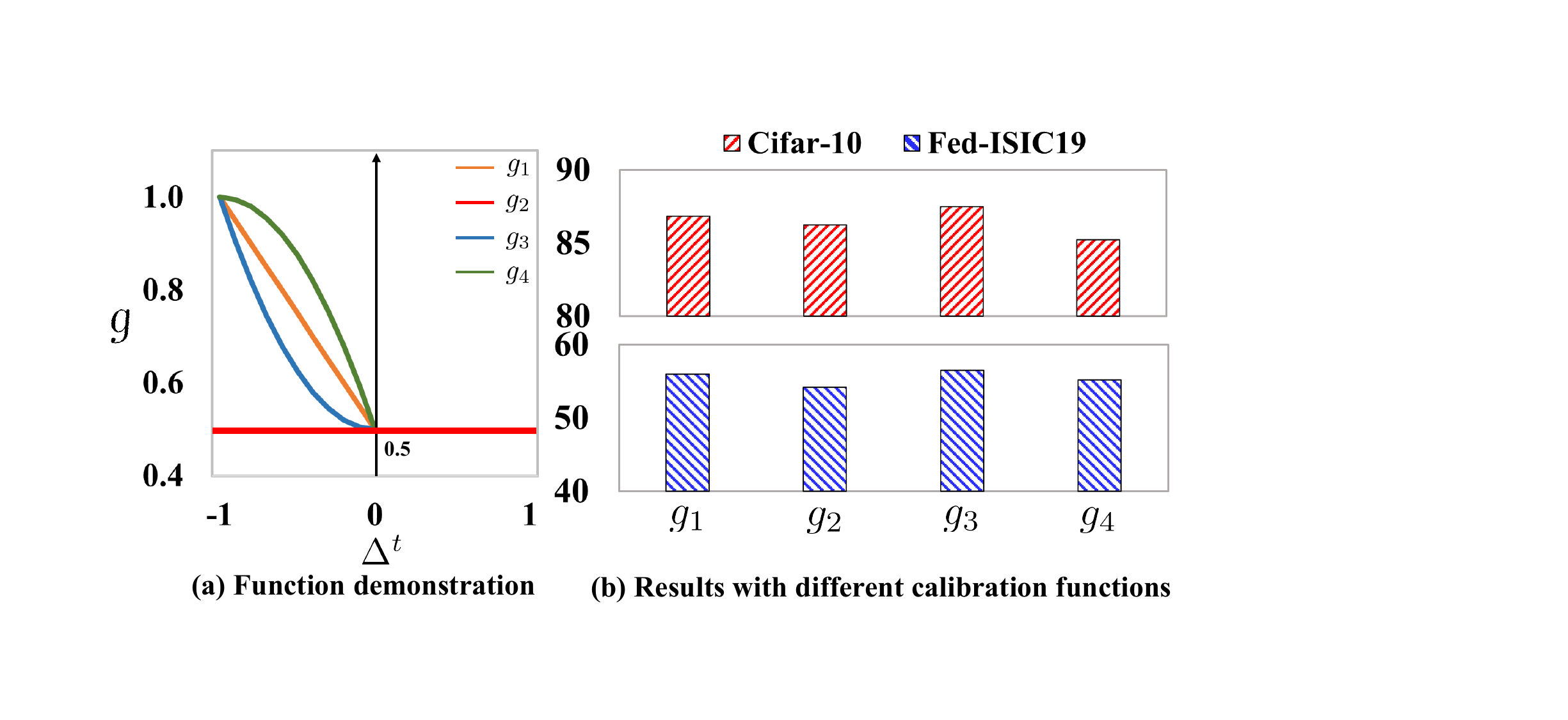}
    \vspace{-0.1in}
    \caption{Calibration function $g(\Delta^t, \lambda)$ study.}
    \label{fig:g_function}
    \vspace{-0.2in}
\end{figure}

Figure~\ref{fig:g_function} (a) and (b) illustrate the $g$ function's representation and its impact on private models respectively. Analyzing the experimental outcomes, we note the following observations: (1) The performance with functions $g_1$, $g_3$, or $g_4$ consistently surpasses that of $g_2$. This suggests that adjusting the value of $\lambda$ in response to accuracy fluctuations positively influences performance on both the Cifar-10 and Fed-ISIC19 datasets. (2) Among $g_1$, $g_3$, and $g_4$, $g_3$ exhibits superior performance on these datasets. A potential explanation is its steeper slope near a $\Delta^t$ value of -1, aiding $g(\Delta^t, \lambda)$ in rapidly declining to identify a more optimal prediction set for the subsequent training epoch. (3) While outcomes with different $g$ function configurations vary, they remain within a stable range, indicating the robustness and adaptability of our proposed \ours to the selection of the $g$ function.

\subsubsection{Alternative Proxy Model Selection}
The proxy model used in all the previous experiments is ResNet-18.
To examine the impact of proxy model selection on performance, we select four widely-used compact models as proxies, including MobileNet-V1~\cite{howard2017mobilenets}, ShuffleNet-V2~\cite{ma2018shufflenet}, and EfficientNet-B0~\cite{tan2019efficientnet}. Model details can be found in Appendix~\ref{apd: model}. The results are shown in Table~\ref{tab:proxy_model_study}, where the aggregation method is still FedAvg. We can observe that the performance variation with these different proxy models generally aligns with their respective sizes and capabilities, which aligns with our expectations. This study of proxy models underscores the resilience of our approach to variations in proxy model choice and further affirms the generalizability of our proposed framework.

\subsubsection{Client Model Architecture Analysis}
In our experiments, we use a mixed combination of client models. To explore the relationship between the model architecture difference and the framework performance, we conduct the following experiment with $\alpha = 0.5$, and the other settings are the same as the experiments shown in Table 1.
In this experiment, we fixed the proxy model, which is ResNet-18, but used different private models. 
{\model-ResNet} denotes all the private client models selected from the ResNet family pool, including ResNet-18, ResNet-34, ResNet-50, RestNet-101, and ResNet 152. 
{\model-VGG} denotes all the private models belonging to the VGG family, including VGG-11, VGG-13, VGG-16, and VGG-19.
{\model-Mix} is used in the main experiments, using the mixed ResNet and VGG client models.

We report the results in Table~\ref{tab:client_model_combination}. We have several exciting observations. First, the private model performs better than the proxy model, which is better than the global model. This observation is the same as we discussed in the main results shown in Table~\ref{tab:hete}. Second, \model-ResNet outperforms \model-Mix, which further outperforms \model-VGG. The observation demonstrates that if the proxy and private models have similar model architectures, the proposed bidirectional knowledge distillation performs the best. 

\begin{table*}[!h]
\centering
\caption{Performamce (\%) of different client model combinations, where all clients' personal models are from ResNet or VGG family.}
\resizebox{0.85\textwidth}{!}{
\begin{tabular}{c|l|ccc|ccc|ccc|ccc} 
\toprule 
\multirow{2}{*}{\textbf{Dataset}} & \multirow{2}{*}{\textbf{Model}} &\multicolumn{3}{c|}{\textbf{FedAvg}}&
\multicolumn{3}{c|}{\textbf{FedProx}} & \multicolumn{3}{c|}{\textbf{pFedMe}} & \multicolumn{3}{c}{\textbf{pFedBayes}}\\\cline{3-14}
&   & global & proxy &private  & global & proxy &private& global & proxy &private & global & proxy &private\\\hline
    \multirow{3}{*}{FMNIST}&
    \model-ResNet&84.55&89.63&91.12&85.51&91.86&93.61&87.41
&93.55&93.64&87.62&92.33&94.86\\
    &\model-Mix  &83.93&89.45&91.22&86.44&91.50
&93.84&87.13&92.05&92.36&87.85&92.11
&93.17\\
&\model-VGG&83.22&88.87&90.86&84.31&91.07&92.46
&86.65&91.77&92.20&87.04&91.98&92.46\\\hline

    \multirow{3}{*}{CIFAR-10}&
\model-ResNet&64.59&83.62&86.88&66.11&83.79&87.25&66.89&84.08&87.10&67.03&
84.55&88.96\\
&\model-Mix&63.39&82.57&86.83&65.86&82.48&86.92&65.22&
83.00&87.24&66.87&84.49&88.67\\
&\model-VGG&61.22&80.61&83.07&64.95&80.87&84.14
&63.40&81.36&84.52&65.48&81.27&86.11\\\hline

    \multirow{3}{*}{CIFAR-100}&
    \model-ResNet&38.95&62.48&65.53&40.66&63.07&
67.89&41.56&62.97&68.16&41.82&63.57&67.94\\
&    \model-Mix&38.17&
61.06&65.69&39.31&61.22&65.45&40.92&62.68&
67.07&41.24&62.96&67.35\\
&\model-VGG&35.96&58.74&62.41&36.77&58.91&62.88
&39.52&59.61&63.05&40.08&61.63&63.85\\\hline

    \multirow{3}{*}{FedISIC-19}&
\model-ResNet&42.86&53.55&55.84&42.84&
55.51&56.90&42.21&53.88&57.03&42.37&55.22&
57.10\\
&\model-Mix&42.30&	53.48&	55.95&
41.77&	54.30&	56.86&
42.35&	53.85&	56.68&
41.82&	54.04&	57.21\\
&\model-VGG&40.76&51.06&52.90&42.04&51.98&53.69
&42.13&52.61&53.12&41.01&52.48&54.55\\
\bottomrule 
\end{tabular}
}
\label{tab:client_model_combination}
\vspace{-0.15in}
\end{table*}

\subsubsection{Upper-bound Performance Exploration}
In this experiment, we aim to investigate the performance upper bound, which will be obtained by training the federated learning framework in the homogeneous setting, where each client uses the largest model (i.e., VGG-19 in our experiments). Since the homogeneous setting aims to train a shared global model, we report the average performance tested on each client with the learned global model from \textbf{Homo-VGG-19} in Table~\ref{tab:app_largest}, compared with the performance of private models obtained by our proposed \ours with $\alpha = 0.5$. To maximize the performance of the private model with our setting, we also test another model, named \textbf{VGG-19+ResNet-18}, in which all clients used VGG-19 as the private model and ResNet-18 as the proxy model. 

\begin{table}[!h]
\centering
\caption{Performance upper-bound analysis.}
\resizebox{0.5\textwidth}{!}
{
\begin{tabular}{c|l|c|c|c|c} 
\toprule 
\textbf{Dataset} & \textbf{Model} & {\textbf{FedAvg}}&
{\textbf{FedProx}} & {\textbf{pFedMe}} & {\textbf{pFedBayes}}\\ \hline

\multirow{3}{*}{{FMNIST}} 
& Homo-VGG-19 &92.86&94.05&94.26&94.87\\
& VGG-19+ResNet-18 & 91.35& 94.12&93.45&94.33\\
&\model &91.22&93.84&92.36&93.17
\\\hline

\multirow{3}{*}{{CIFAR-10}} 
& Homo-VGG-19 & 87.89&88.62&89.10&89.92\\
& VGG-19+ResNet-18 & 87.04 & 87.80& 87.98&89.55\\
&\model&86.83&86.92&87.24&88.67\\\hline

\multirow{3}{*}{{CIFAR-100}} 
& Homo-VGG-19 & 65.91&67.52&69.81&69.98\\
& VGG-19+ResNet-18 & 65.77&66.71&67.89&68.89\\
& \model &65.69&65.45&67.07&67.35\\\hline

\multirow{3}{*}{{FedISIC-19}} 
& Homo-VGG-19 & 56.89 & 58.94 & 59.06 & 59.22\\
& VGG-19+ResNet-18 & 56.62&57.12&57.59&57.96\\
&\model&55.95&56.86&56.68&57.21\\
\bottomrule 
\end{tabular}
}
\label{tab:app_largest}
\vspace{-0.3in}
\end{table}

We can observe that the upper bound performance obtained by the Homo-VGG-19 model performs the best, and VGG-19+ResNet-18 is better than \ours but with marginal improvements, which aligns with our expectations. 
However, our model and VGG-19+ResNet-18 use ResNet-18 (with 11.17 million parameters) as the proxy model to exchange information between clients and the server. However, the Homo-VGG-19 model has 144 million parameters, which is 12.9X of our model. Considering the communication costs, \ours is an effective solution to address the model heterogeneity challenge in federated learning.








\subsubsection{Scalability Analysis}
In this experiment, we validate the scalability of the proposed \ours with a different number of clients on the CIRFA-10 dataset with a sample rate 10\% and $\alpha = 0.5$. The client model pool contains ResNet-18, ResNet-34, and ResNet-50. The proxy model used in this experiment is MobileNet-V1, and the aggregation method is FedAvg.
Table~\ref{tab:scale} presents our experimental findings with client counts of 50, 100, and 300. A noticeable trend is the marginal decline in performance as the number of clients rises. This outcome is attributable to the distribution of the same total volume of training data among a greater number of clients, resulting in reduced data availability per client. Such a scenario leads to a performance dip within a specific range. These results effectively illustrate the scalability of our proposed \model, highlighting its adaptability to varying client numbers in federated learning environments.

\subsection{Results of the Homogeneous Model Setting}\label{sec:homo}
\begin{figure}[t]
\label{fig:homo}
    \centering
    \includegraphics[width=0.43\textwidth]{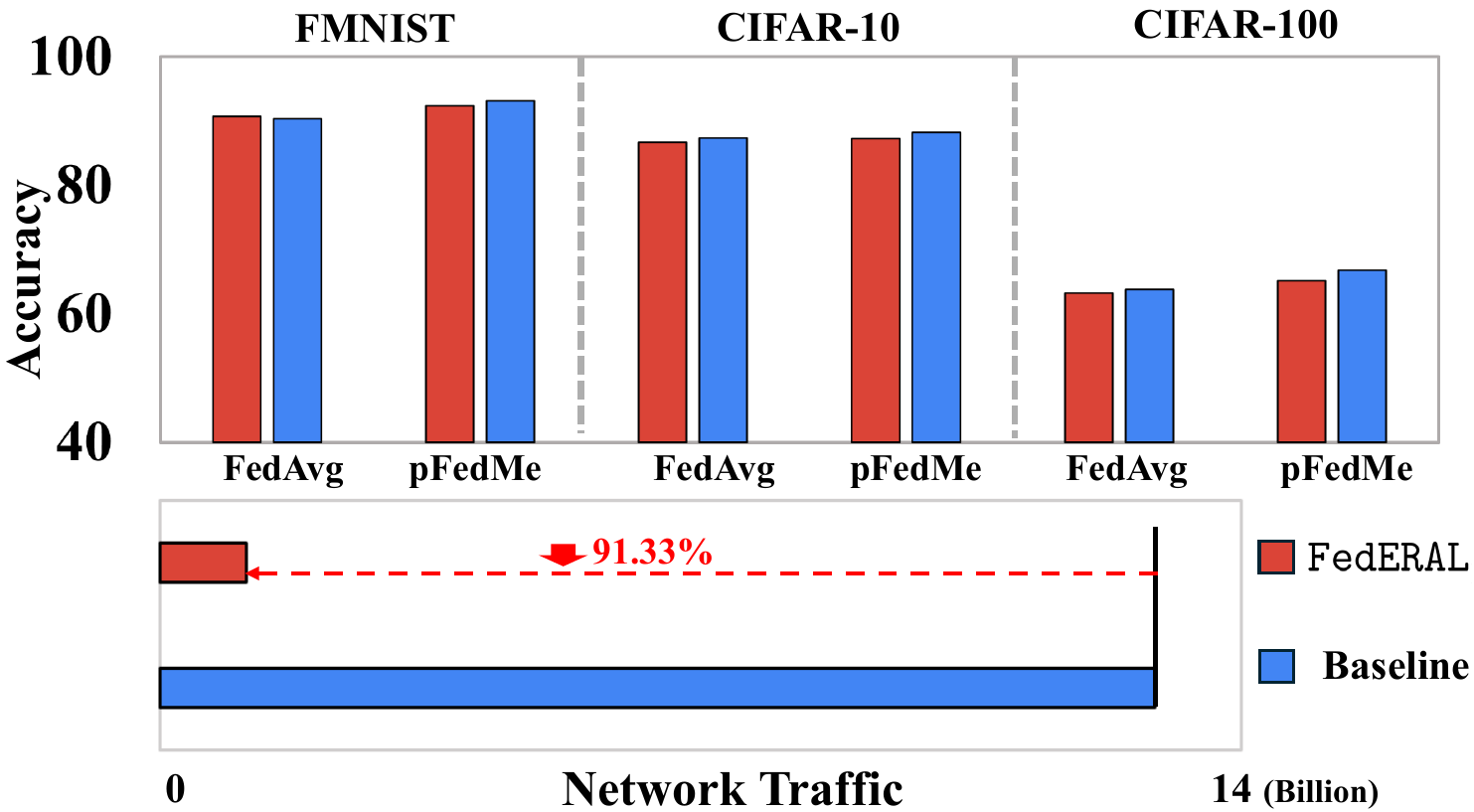}
    \vspace{-0.15in}
        \caption{Homogenenous evaluation.}
     \vspace{-0.2in}
\end{figure}

While the primary focus of the proposed \ours is addressing the challenge of model heterogeneity, it showcases versatility by extending its application to the model homogeneous setting. To illustrate its generalization ability, we homogenize all clients to utilize the same model structure, VGG-11, with \ours employing ResNet-18 as the proxy model in this experiment. A comparison is made with two representative federated learning approaches, FedAvg and pFedMe, and the results are outlined in Figure~\ref{fig:homo}. Notably, \ours, with fewer parameters (only around 8\% of baselines' parameters), achieves comparable performance with existing homogeneous federated learning models. This underscores the effectiveness and advantages of \ours, even in a homogeneous model setting. \textbf{\textit{More experimental results can be found in Appendix Sections~\ref{apd:converge} to~\ref{apd:resource}}}.

\begin{table}[t]
\centering
\caption{scalability study in the cross-device setting.}
{
\begin{tabular}{l|ccc} 
\toprule 
\textbf{Setting}  & 50& 100 &300\\
\midrule
\model$_{\text{global}}$&62.88&59.40&56.01\\
\model$_{\text{proxy}}$&84.33&81.46&77.90\\
\model$_{\text{private}}$&86.74  &84.26&80.15\\ 

\bottomrule 
\end{tabular}
}
\label{tab:scale}
\end{table}

%% file: section/conclusion.tex
\vspace{-0.1in}
\section{Conclusion}
We have designed \ours to effectively tackle the challenges of model heterogeneity without relying on public data. By implementing a novel uncertainty-based asymmetrical reciprocity learning method, we have not only demonstrated the feasibility but also showed the superiority of \ours in handling diverse model structures while safeguarding client privacy and minimizing communication costs. Our comprehensive experiments across multiple benchmark datasets illustrate the effectiveness of \ours in different scenarios. We believe that \ours significantly contributes to the advancement of general federated learning and holds immense potential for practical applications in real-world scenarios.

\section*{Acknowledgements} 
This work is partially supported by the National Science Foundation under Grant No. 2333790, 2212323, and 2238275 and the National Institutes of Health under Grant No. R01AG077016.

\section*{Impact Statement}

The introduction of \ours represents a significant advancement in the field of federated learning (FL), addressing crucial challenges associated with heterogeneous model aggregation. By leveraging small identical proxy models, \ours ensures secure and efficient information exchange among clients. The implications of this research are far-reaching, promising to enhance the application of federated learning in real-world scenarios where data privacy and efficient communication are paramount. By eliminating the reliance on public data and ensuring robust performance across diverse conditions, \ours has the potential to revolutionize the deployment of FL in sensitive and resource-constrained environments, including healthcare, finance, and beyond.

The proposed framework has two major limitations. First, selecting the proxy model is non-trivial and further influences the framework performance. In future work, we plan to design an automatic strategy to adaptively select the proxy model according to the types of private models. Second, although the proposed framework is effective and efficient, the designed uncertainty-based asymmetrical reciprocity learning between the proxy and private model runs slightly slower than the naive bidirectional knowledge distillation. Thus, we need to develop a more efficient approach to accelerate this step. From our perspective, there is no negative social impact on the designed framework.

%% file: section/appendix.tex
\newpage
\appendix
\onecolumn
\section{Algorithm Flow}\label{apd:algorithm}
In this subsection, we show the whole algorithm flow of our proposed \ours in Algorithm~\ref{alg:alg_flow}.


\begin{algorithm*}[ht]
\small
\SetAlgoLined
\DontPrintSemicolon
\Input{ Client training data ${\mathcal{D}_1, \cdots,\mathcal{D}_N}$, private models $\{\mathbf{w}_1,\cdots,\mathbf{w}_N$\}, proxy models  $\{\mathbf{p}_1,\cdots,\mathbf{p}_N\}$, validation data $\{\mathcal{D}_1^\prime,\cdots,\mathcal{D}_N^\prime\}$, communication round T, local training epoch $R$, and hyperparameters.} 
\For{\textup{each communication round }$t = 1,2,\cdots$,T}{

    \kwClient{}{
        \For{\textup{each epoch} $r = 1, \cdots, R$}{

        \For{\textup{active client} $i \in [1, \cdots, B]$}{ 

        Conduct uncertainty-based asymmetrical reciprocity learning following the Algorithm~\ref{alg:uarl}. 
        }}
        Obtain private models $\{\mathbf{w}^{t}_1,\cdots,\mathbf{w}^{t}_N\}$ and proxy models  $\{\mathbf{p}^{t}_1,\cdots,\mathbf{p}^{t}_N\}$;

        Upload  proxy models  $\{\mathbf{p}^{t}_1,\cdots,\mathbf{p}^{t}_N\}$ to the server;
    
}

    \kwServer{}{
        \textup{Obtain the aggregated model $\mathbf{p}^t_g$ via FedAvg or other existing methods;}
        
        \textup{Distribute the aggregated model $\mathbf{p}^g_t$ back to clients.}
    }
    }

\caption{Algorithm Flow of \model.}
\label{alg:alg_flow}
\end{algorithm*}


\section{Details of Conformal Prediction}\label{apd:conformal}

\subsection{Conformal Prediction}

Conformal prediction is a promising statistical framework for describing the prediction uncertainty~\cite{vovk2005algorithmic, shafer2008tutorial,angelopoulos2021gentle,angelopoulos2022conformal,barber2023conformal,sharma2023pac}. It guarantees finite-sample coverage under the mild assumption of exchangeability. The main idea of conformal prediction involves establishing a non-conformity score $S$ to assess the conformity between new test data and existing data distribution. Compared with existing traditional methods for estimating prediction uncertainty, such as Bayesian neural networks~\cite{neal2012bayesian,kuleshov2018accurate,maddox2019simple} and Platt scaling~\cite{platt1999probabilistic,guo2017calibration}, conformal prediction offers numerous advantages, including its post-hoc nature, and being distribution-free and model-agnostic.

The original version of conformal prediction, also known as full conformal prediction or transductive conformal prediction, is regarded for its exceptional uncertainty estimation capabilities. 
However, its considerable computational cost restricts its practical application. To address this, several alternative methods have been developed to reduce the computational demands of full conformal prediction while retaining its core utility, for example, Split Conformal Prediction (Inductive CP)~\cite{papadopoulos2002inductive}, Cross-CP~\cite{vovk2015cross}, and jackknife+~\cite{barber2021predictive}. In particular, Split Conformal Prediction has garnered significant attention due to its ease of implementation. 
Our paper is based on split conformal prediction, which we will introduce below.

\subsection{Split Conformal Prediction Model Training}

Next, we describe the process for deriving $\mathbf{cp}_{i}^{t}$ as outlined in Eq.~\eqref{eq:conformal_model} following the methodology presented in the work~\cite{angelopoulos2020uncertainty}. We describe the process of obtaining $\mathbf{cp}_i^t$ in Algorithm~\ref{alg:alg_flow_cmodel} below.

\begin{algorithm*}[ht]
\small
\SetAlgoLined
\DontPrintSemicolon
\Input{A trained proxy model $\mathbf{p}_i^t$ using the training dataset $\mathcal{D}_{i}$, validation data $\mathcal{D}_i^\prime$.}

\begin{itemize}
    \item \textbf{Step 1:} Define the nonconformity score function $S$ based on the nonconformity measure and given hyperparameters;    
    \item \textbf{Step 2:} Calculate the nonconformity score $s_j$ for each $\mathbf{x}_j \in \mathcal{D}_i^{\prime}$, i.e., $s_j=S(\mathbf{x}_j)$, forming the nonconformity score set $\{s_j\}$;
    \item \textbf{Step 3:} Obtain the score quantile $Q_{\theta}$ related to $\theta$ on the nonconformity score set, i.e., $Q_{\theta}:=\frac{1}{n}\lceil(1-\theta)(n+1)\rceil$-th quantile \\ of $\{s_j\}_{j=1}^n$, where $n  = |\mathcal{D}_{i}^{'}|$;
    \item \textbf{Step 4:} Integrate $\mathbf{p}_{i}^t$ with $Q_{\theta}$. $\mathbf{cp}_{i}^t$ is first initialized by trained $\mathbf{p}_{i}^t$ and then predicts and calculates the non-conformity score \\ on the test data coupled with the candidate label, which is then compared with the $Q_{\theta}$.     
\end{itemize}
\Return{The trained conformal model $\mathbf{cp}_{i}^t$.}
\caption{Algorithm Flow of Cmodel Training with Eq.~\eqref{eq:conformal_model}.}
\label{alg:alg_flow_cmodel}
\end{algorithm*}

\subsection{Regularized Adaptive Prediction Sets Converge Guarantee}

Note that our proposed dynamic adjustment mechanism retains all the characteristics of traditional conformal prediction requirements. Additionally, it ensures the isolation of $\mathcal{D}_{i}$ and $\mathcal{D}^{'}_{i}$ throughout the entire federated learning process, including during local training phases. Thus, regularized adaptive prediction sets converge can be guaranteed. More details about the \textbf{upper bound proof} and \textbf{lower bound proof} can be found in Section D in~\cite{angelopoulos2021gentle}.

\section{Datasets}\label{apd:datasets}
We assess the effectiveness of the proposed \ours approach through image classification tasks conducted in both cross-device and cross-silo scenarios. For the \textbf{cross-device} evaluation, we employ three image classification datasets -- FMNIST, CIFAR-10, and CIFAR-100. To introduce heterogeneity in federated learning, a methodology inspired by existing work~\cite{yurochkin2019bayesian,hsu2019measuring} is followed. This involves manipulating the concentration parameter $\alpha$ of the Dirichlet distribution to partition the datasets. Here, $\alpha$ signifies the label heterogeneity across clients, with smaller values concentrating labels on a few categories for a client (where $\alpha \to 0$ implies a client stores data with a single category). Conversely, larger $\alpha$ values lead to clients holding a more diverse set of label categories (where $\alpha \to +\infty$ indicates each client possesses data spanning all categories). Our data distribution process involves initially allocating data to clients, followed by a subsequent split into local model training, testing, and conformal learning sets in a 7:2:1 ratio.  \emph{While the splitting method guarantees an identical distribution for both sets, aligning effectively with the personalized federated learning setting, it is noteworthy that the increased value of $\alpha$ introduces heightened difficulty in the classification tasks.}

For the \textbf{cross-silo} setting, the Fed-ISIC19 dataset~\cite{ogier2022flamby} is employed, featuring 23,247 dermoscopy images encompassing eight distinct types of melanoma. Following the data partition strategy of FLamby~\cite{ogier2022flamby}, the number of training/testing data on six clients is distributed as 9,930/2,483, 3,163/791, 2,690/673, 655/164, 351/88, and 180/45, respectively. The cross-silo setting necessitates the involvement of all clients in the training process at each communication round.

\section{Baseline Introduction}\label{apd:baseline}
In our paper, though there is no previous heterogeneous FL work sharing the exact setting with our work, we introduce the selected baselines used for the homogeneous setting comparison. 
\begin{itemize}
    \item \textbf{FedAvg} \cite{mcmahan2017communication}: It is the vanilla baseline. In this approach, active local clients train their models and send the parameters to a central server. The server then computes the average of these local model parameters and redistributes the aggregated global model to the active clients for subsequent rounds of local training.

   \item \textbf{FedProx}~\cite{li2020federated}: this work introduces a proximal term in the local training of each client, quantifying the divergence between the local and global models. This term acts as a constraint, ensuring that the local models' personalized optimization does not deviate significantly from the global model.
    \item \textbf{pFedMe}~\cite{t2020personalized}: It uses regularized loss and decouples the personalization problem into a bi-level optimization.  pFedMe aims to develop personalized models for each client. It does so by integrating Moreau envelopes into the learning process to help regularize the local updates during training;
    \item \textbf{pFedBayes}~\cite{zhang2022personalized}: It proposes an algorithm to take consideration of the global distribution while conducting local model training. By leveraging Bayesian inference, pFedBayes not only customizes models to fit individual client needs better but also provides a robust framework for managing the inherent uncertainties and variabilities in distributed datasets.
\end{itemize}
\emph{Notably, these approaches are all used as model aggregation solutions in the experiments due to the flexibility of our proposed framework.}

\section{Model Details and Implementation}\label{apd: model}
We introduce the models that we use in our experiment one by one. For all the following \textbf{client private models}, we adjust the dimension of the linear layer to fit the number of classes of the datasets accordingly.
\begin{itemize}
    \item \textbf{ResNet family}: ResNet~\cite{he2016deep} is a type of convolutional neural network (CNN) architecture proposed in 2015. The key innovation of ResNet is its use of residual blocks. These blocks allow the network to learn residual functions with reference to the layer inputs, instead of learning unreferenced functions. In our work, we use ResNet-18, 34, 50, 101, and 152. The implementation is based on the Pytorch official library\footnote{\url{https://pytorch.org/vision/main/models/resnet.html}}.
    
    \item \textbf{VGG family}: VGG~\cite{simonyan2014very}, short for Visual Geometry Group, refers to a deep CNN architecture. VGG was one of the first to demonstrate that depth of the network (i.e., the number of layers) is a critical component for achieving high performance in visual recognition tasks. In our work, we use VGG-11, 13, 16, and 19. The implementation is based on the Pytorch official library\footnote{\url{https://pytorch.org/hub/pytorch_vision_vgg/}}.
\end{itemize}

Except for ResNet-18, we also test the following models as the \textbf{proxy modes} in our experiments:
\begin{itemize}   
    \item \textbf{ShuffleNet}: ShuffleNet\cite{zhang2018shufflenet} is an efficient CNN designed primarily for mobile and computing devices with limited computational capacity. The key innovation in ShuffleNet is the use of pointwise group convolutions and channel shuffle operations to significantly reduce the computational cost and the number of parameters. In our work, we use ShuffleNet V2~\cite{ma2018shufflenet} and implement it using the Pytorch library\footnote{\url{https://pytorch.org/hub/pytorch_vision_shufflenet_v2/}}.
    \item \textbf{MobileNet}: MobileNet~\cite{howard2017mobilenets} is a class of efficient CNN architectures designed specifically for mobile and embedded vision applications. The key innovation in MobileNet is the use of depthwise separable convolutions, which significantly reduce the number of parameters and the computational burden compared to standard convolutions used in more traditional CNN architectures. In our experiment, we use MobileNet V1~\cite{howard2017mobilenets} and implement it based on the Github resource \footnote{\url{https://github.com/jmjeon94/MobileNet-Pytorch/blob/master/MobileNetV1.py}}.
    \item \textbf{EffcientNet}: EfficientNet is an efficient CNN structure introduced in ~\cite{tan2019efficientnet}. The primary innovation of EfficientNet lies in its novel compound scaling method, which uniformly scales the depth, width, and resolution of the network with a set of fixed scaling coefficients. In our paper, we use EfficientNet-B0 and implement via Pytorch library\footnote{\url{https://pytorch.org/vision/main/models/efficientnet.html}}.
\end{itemize}

\section{Hyperparameter Details}\label{apd:implementation}
Our experimental setup involves 100 communication rounds, 100 clients, a 20\% sample ratio for the cross-device experiments, and five local training epochs. 
All experiments are conducted on an NVIDIA A100 with CUDA version 12.0, running on a Ubuntu 20.04.6 LTS server. All baselines and the proposed \ours are implemented using PyTorch 2.0.1. 
For the local update, we set the learning rate as 0.0001, the batch size is 16, and the optimizer used in the optimization is Adam. Following the provided value in the work~\cite{angelopoulos2020uncertainty}, we set $\lambda = 0.5$, $\kappa_{reg} = 5$, and 
 $\theta = 0.1$ for the local conformal model and proxy conformal model in Eq.~\eqref{eq:cp}, the conformal learning batch size is 32. The learning rate in the Platt scaling process is 0.01, and the maximum iteration is 10 following the default setting.

\section{Empirical Study of Convergence}\label{apd:converge}
\begin{figure*}[h!]
\centering
\includegraphics[width=0.8\textwidth]{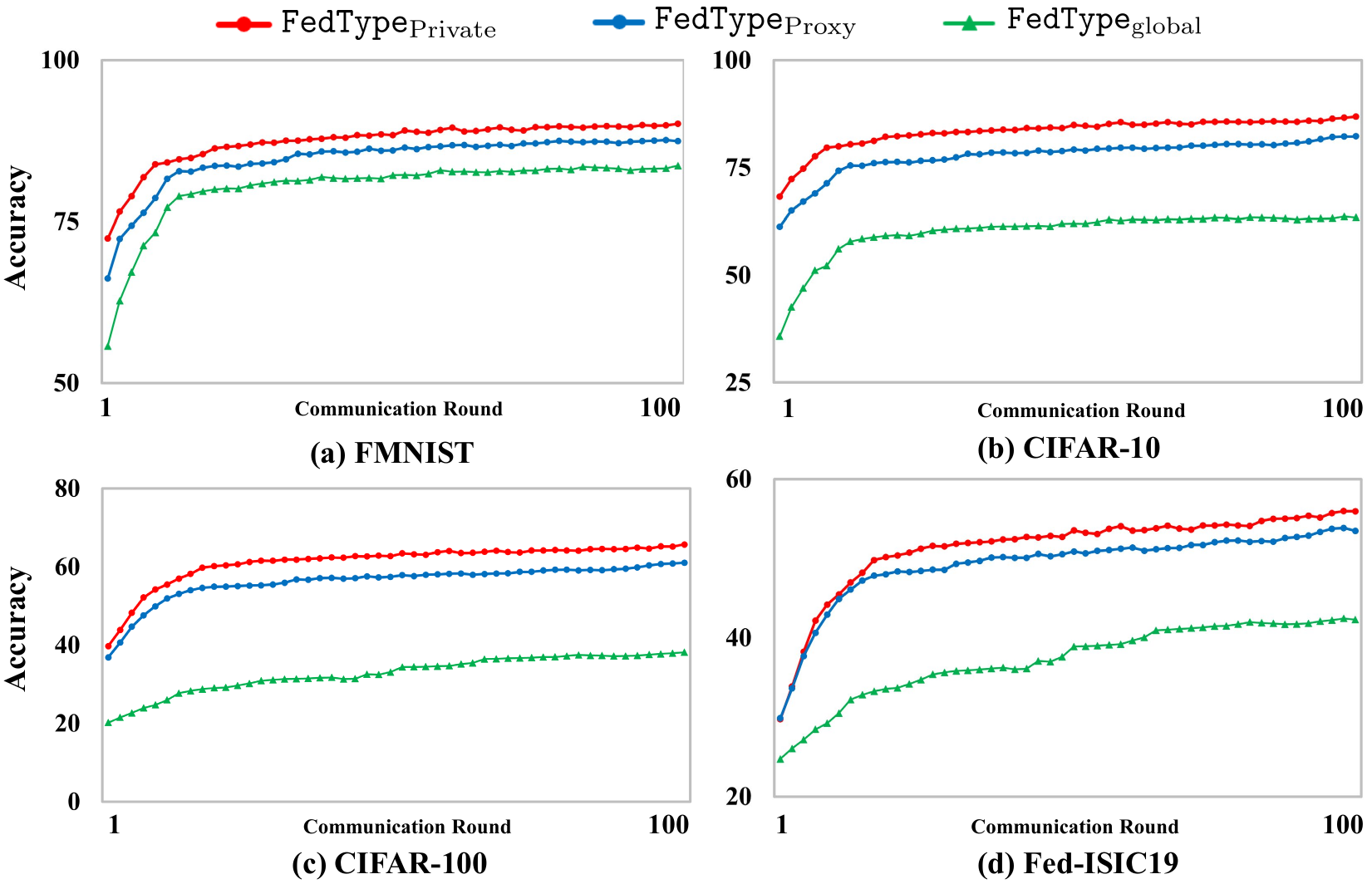}
    \caption{The empirical convergence of our proposed \ours. }
    \label{fig:app_converge}
\end{figure*}

In this section, we present the convergence results of our proposed \ours framework as illustrated in Figure~\ref{fig:app_converge} across different datasets: FMNIST, CIFAR-10, CIFAR-100, and Fed-ISIC19, shown in panels (a), (b), (c), and (d),  respectively. These results are obtained under the setting of $\alpha = 0.5$ with FedAvg employed as the aggregation method. The x-axis represents the number of communication rounds, while the y-axis tracks the average accuracy of clients. Observations from these results indicate that the private model, the proxy model, and the global model within the \ours framework all achieve convergence. This effectively demonstrates the robust convergence properties of our proposed approach empirically.

\section{Study of Consensus Weight $\eta$}
\begin{figure*}[h!]
\centering
\includegraphics[width=0.8\textwidth]{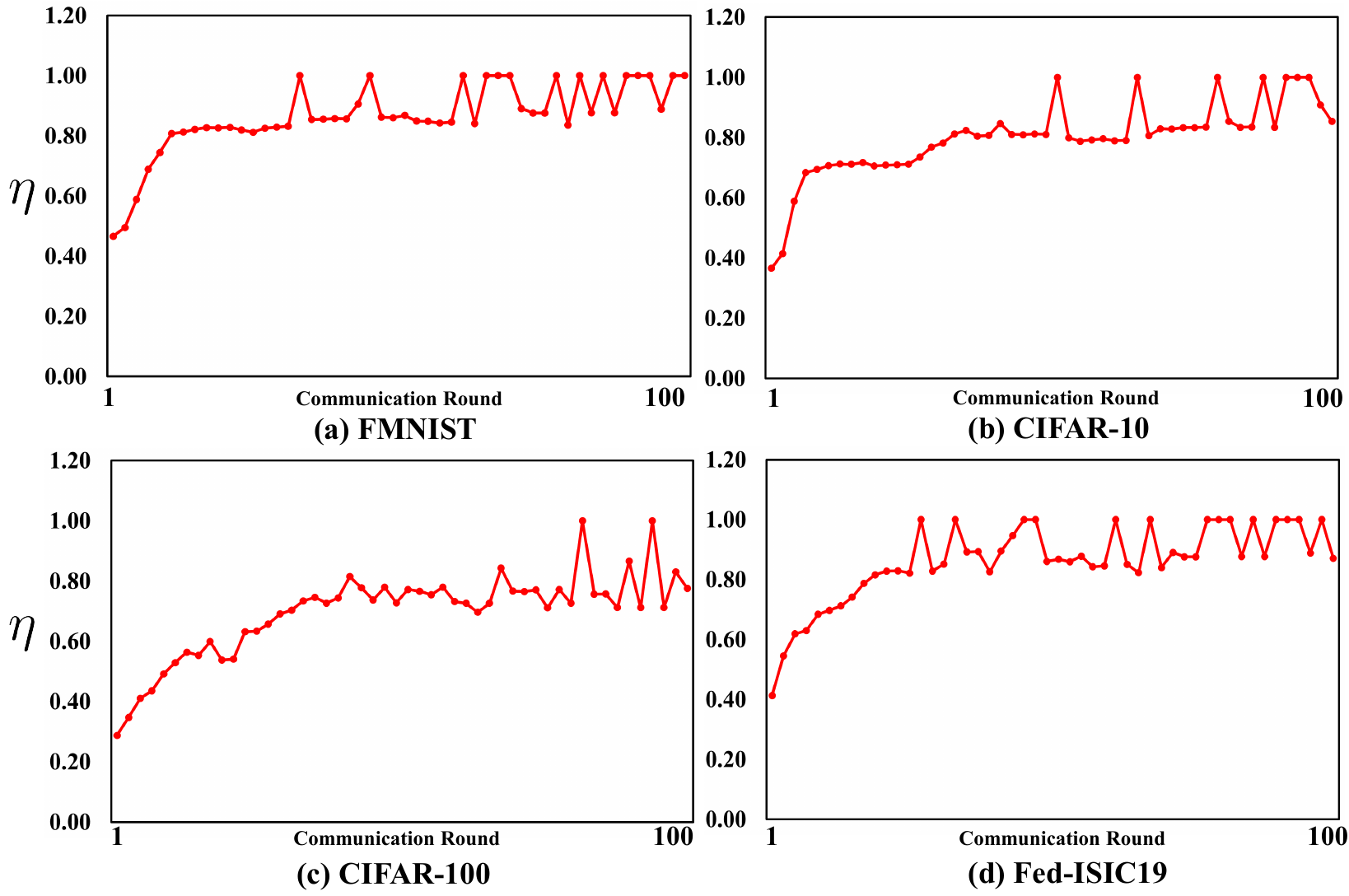}
    \caption{The value of $\eta$ of our proposed \ours. }
    \label{fig:app_eta}
\end{figure*}

As shown in Section~\ref{sec:weight} and Eq.~\eqref{eq:certainty_KD}, 
$\eta$ serves as a crucial metric in our framework, quantifying the degree of consensus to regulate knowledge transfer from the proxy model $\mathbf{p}_{i}^{t}$ to the larger model $\mathbf{w}_i^{t}$. In this section, we analyze the behavior of $\eta$ across various datasets: FMNIST, CIFAR-10, CIFAR-100, and Fed-ISIC19, as depicted in Figure~\ref{fig:app_eta} in panels (a), (b), (c), and (d), respectively, where x-axis denotes the number of communication rounds, and y-axis represents the average $\eta$ values of all training data for all active/selected clients at round $t$. These observations are made under the setting of $\alpha = 0.5$ for FMNIST, CIFAR-10, and CIFAR-100, utilizing FedAvg as the aggregation method. 

The results indicate a consistent increase in the value of $\eta$ across communication rounds, suggesting that the large model and the proxy model are achieving greater consensus as training progresses. Notably, an $\eta$ value of 1 implies complete uniformity in the uncertainty set for input data between the large and proxy models. Interestingly, for simpler datasets like FMNIST and Fed-ISIC19, $\eta$ attains more instances of 1 and converges more rapidly compared to the more complex CIFAR-10 and CIFAR-100 datasets. These experimental findings highlight the dynamic nature of $\eta$ throughout the iterative learning process and validate our strategy of assigning increased weight to the knowledge transferred from the proxy model to the large model as consensus grows.

\section{Calibration Function Study}

In Section~\ref{sec:dcp}, we detailed a dynamic calibration function designed to adjust the prediction set size in response to changes in model performance. This section delves deeper into the training process, specifically examining changes quantified with $\alpha = 0.5$ and FedAvg as the aggregation method. We focus on the FMNIST, CIFAR-10, CIFAR-100 datasets under a cross-device setting, and Fed-ISIC19 under a cross-silo setting. At the 50th communication round, we highlight changes in active clients' \textbf{average proxy set size} and \textbf{average client accuracy} (last batch of the training epoch), as presented in Table~\ref{tab:app_calibration}.

From these results, we draw several insights: (1) In scenarios where accuracy improves across training epochs, such as with CIFAR-10 and Fed-ISIC19, the prediction set size of the proxy model remains fairly consistent. This stability is attributed to our specifically designed function $g(\Delta^t, \lambda)$. In instances where $\Delta^t$ falls within the range of $(0, 1]$, the function maintains $g(\Delta^t, \lambda) = \lambda$, with $\lambda$ consistently set at 0.5. (2) Conversely, where there's a decrease in accuracy, as observed in the CIFAR-100 dataset between the 1st and 3rd epochs (accuracy dropping from 62.15\% to 61.42\%), the prediction set size correspondingly reduces from 3.06 to 2.56. In contrast, between the 3rd and 5th epochs of the same dataset, as accuracy increases from 52.56\% to 52.77\%, the set size also rises from 2.63\% to 2.69\%. Similar patterns are noted in the FMNIST dataset. These findings underscore the efficacy of our dynamic calibration function in adapting the prediction set size in tandem with fluctuations in model accuracy during the training process.

\begin{table*}[!h]
\centering
\caption{Calibration function quantitative analysis. $|\mathcal{S}^{50}|$ denotes the average size of the prediction sets by proxy models.}
 \resizebox{0.95\textwidth}{!}
{
\begin{tabular}{c|cc|cc|cc|cc} 
\toprule 

\textbf{Dataset} 
&  \multicolumn{2}{c|}{\textbf{FMNIST}}&
\multicolumn{2}{c|}{\textbf{CIFAR-10}} & \multicolumn{2}{c|}{\textbf{CIFAR-100}}&\multicolumn{2}{c}{\textbf{Fed-ISIC19}}\\\hline
\textbf{Epoch}&Accuracy&$|\mathcal{S}^{50}|$&Accuracy&$|\mathcal{S}^{50}|$&Accuracy&$|\mathcal{S}^{50}|$&Accuracy&$|\mathcal{S}^{50}|$\\\hline
1&87.85 (-)&2.15 (-)&83.26 (-)&2.13 (-)&62.15 (-)&3.06 (-)&52.10 (-)&2.69 (-)\\
2&87.62 ($\downarrow$)&1.75 ($\downarrow$)&{83.57 ($\uparrow$)}&{2.11} ($\downarrow$)&61.52 ($\downarrow$)&2.75 ($\downarrow$)&{52.47 ($\uparrow$)}&{2.56 ($\downarrow$)}\\
3&88.21 ($\uparrow$)&2.00 ($\uparrow$)&83.92 ($\uparrow$)&2.13 ($\uparrow$)&61.42 ($\downarrow$)&2.56 ($\downarrow$)&52.56 ($\uparrow$)&2.63 ($\uparrow$)\\
4&88.43 ($\uparrow$)&2.19 ($\uparrow$)&84.11 ($\uparrow$)&2.13 (-)&62.56 ($\uparrow$)&2.62 ($\uparrow$)&{52.62 ($\uparrow$)} &{2.56($\downarrow$)}\\
5&88.56 ($\uparrow$) &2.25 ($\uparrow$)&84.21 ($\uparrow$)&2.19 ($\uparrow$)&62.87 ($\uparrow$)&3.12 ($\uparrow$)&52.77 ($\uparrow$) &2.69 ($\uparrow$)\\
\bottomrule 
\end{tabular}
}
\label{tab:app_calibration}
\end{table*}

\begin{figure*}[t]
  \centering
  \includegraphics[width=0.7\textwidth]{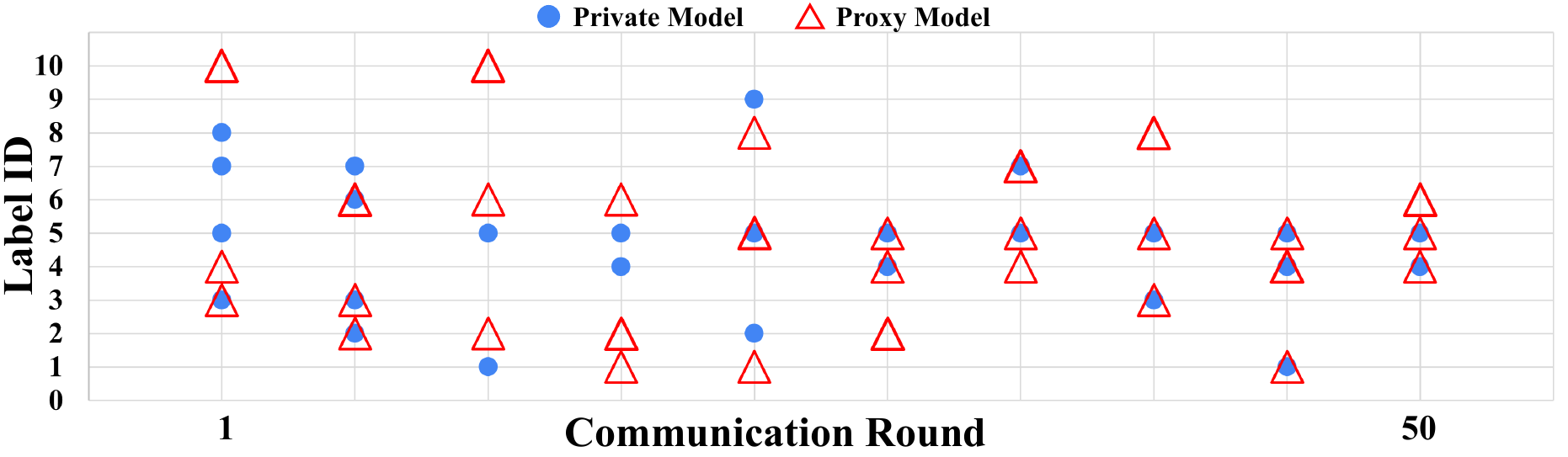}
  \vspace{-0.15in}
  \caption{Prediction set consensus study.}
  \vspace{-0.15in}
  \label{fig:app_label}
\end{figure*}

\section{Conformal Prediction Set Visualization}

In this section, our focus is on evaluating the performance of conformal models for both private and proxy models on identical data points throughout the iterative training process. This evaluation is aimed at understanding how these models progress towards consensus on a certain data sample $\mathbf{x}_j$. Utilizing the CIFAR-10 dataset and adhering to the settings outlined in Table~\ref{tab:hete}, we graphically represent the prediction set of one data sample randomly selected from CIFAR-10 for both private and proxy models in Figure~\ref{fig:app_label}. On this graph, the x-axis represents the communication round, while the y-axis corresponds to label IDs (ranging from 1 to 10, corresponding to the dataset's labels).

From our analysis, it is evident that the labels within the prediction sets generated by the conformal models of both private and proxy models increasingly overlap as the communication rounds progress. Notably, in later communication rounds, we observe a reduction in the size of both prediction sets, with each set encapsulating the ground truth label 5. This convergence of label predictions between the two models' prediction sets suggests a gradual move toward consensus within a specific range for the given input data. This observation is crucial as it demonstrates the effectiveness of the iterative training process in aligning the outputs of the private and proxy models, contributing significantly to the overall model consensus and performance.

\section{Experiment Results on Text Classification Task}
In addition to our previous experiments, we also conducted tests on text classification using the AG news dataset to further validate our \ours framework. For these experiments, we created a heterogeneous model pool by adding 3, 4, or 5 linear layers to the base of a pre-trained DistillBERT model. Specifically, we used DistillBERT with 3 linear layers as the proxy model in the heterogeneous setting and DistillBERT with 4 linear layers for the homogeneous setting. The experimental setup included using $\alpha = 0.5$ and FedAvg as the aggregation method, with a total of 50 communication rounds, keeping all other settings consistent with those in Table~\ref{tab:hete}. During training, we only fine-tuned the appended linear layers, keeping the pre-trained DistillBERT layers fixed. The results, presented in Table~\ref{tab:app_text}, affirm the effectiveness of \ours in text classification tasks.

Interestingly, despite the utilization of a common pre-trained language model, which led to similar accuracy levels as observed in our image classification tasks, there were still noticeable differences in the performance of the global, proxy, and private models. This aligns with the observations from our image classification experiments, thereby reinforcing the efficacy of \ours in handling text classification tasks. These results not only demonstrate the versatility of our approach but also its adaptability across different types of data.
\begin{table*}[t]
\centering
\caption{Performance (\%) of the text classification task under the heterogeneous cross-device settings .}
 \resizebox{0.7\textwidth}{!}
{
\begin{tabular}{l|c|c|c} 
\toprule 

\textbf{Dataset}& \model$_{\text{global}}$&
\model$_{\text{proxy}}$ & \model$_{\text{private}}$\\
\midrule
Heterogenous&91.21&91.86&92.16\\
Homogenous&90.13&90.89&91.44\\
\bottomrule 
\end{tabular}
}
\label{tab:app_text}
\end{table*}

\section{Homogeneous Results on Fed-ISIC19 dataset}

\begin{wrapfigure}{r}{0.4\textwidth}
\vspace{-0.3in}
  \centering
  \includegraphics[width=0.35\textwidth]{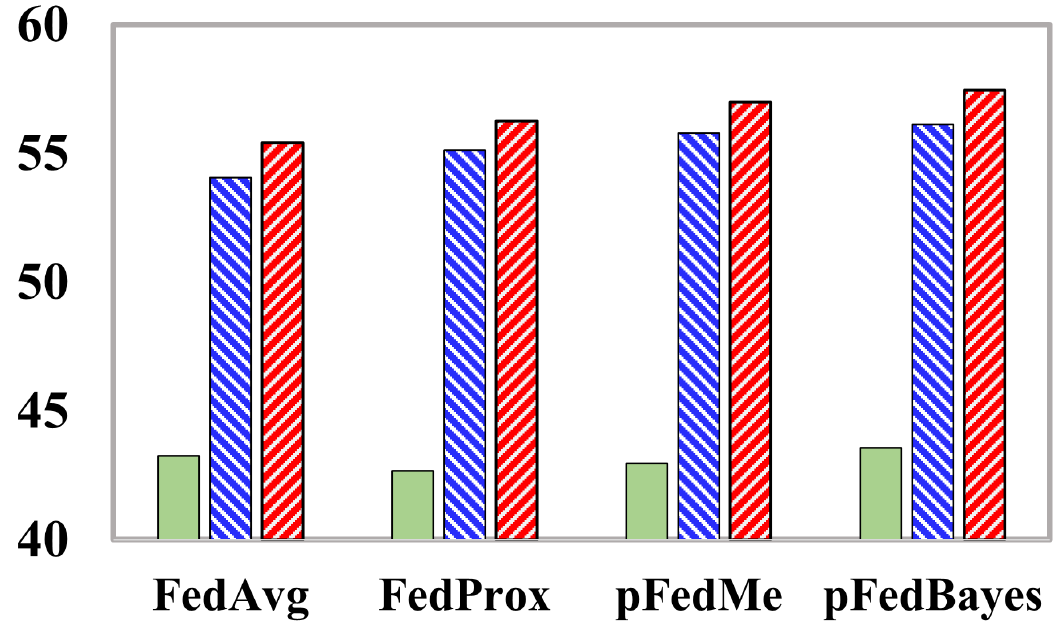}
  \vspace{-0.1in}
  \caption{Homogeneous evaluation with Fed-ISIC19 dataset under the cross-silo setting.}
  \vspace{-0.15in}
  \label{fig:app_homomed}
\end{wrapfigure}

In this section, we report the experiment results on Fed-ISIC19 under the homogeneous setting due to the page limit of the main paper. We maintain all the other settings as the heterogeneous setting except replacing the private model of clients with VGG-11, which is the same as the setting in Section~\ref{sec:homo}. The results are shown in figure~\ref{fig:app_homomed}. Based on the experiment results, we observe that they generally align with the results under the heterogenous setting in Figure~\ref{fig:cross-silo}. It further demonstrates the effectiveness and robustness of our proposed \ours under both the heterogenous setting and the homogeneous setting. 

\begin{table*}[!h]
\centering
\caption{Comparision of the average of 3 epoch training time (second) on datasets}
\resizebox{0.9\textwidth}{!}{
\begin{tabular}{lcccc}
\toprule
Setting  & FMNIST&CIFAR-10&CIFAR-100&Fed-ISIC19\\
\midrule
Private model training in \ours &265.67&279.33&312.67&526.00\\
Private model training with symmetrical KD loss &238.00&241.67&275.33&493.67\\\hline
Proxy model training in \ours&161.33&164.67&176.33&371.00\\
Proxy model training with only CE loss&129.67&131.33&152.00&253.33\\
\bottomrule
\end{tabular}}
\label{tab:app_time}
\end{table*}

\section{Resource Usage Discussion}\label{apd:resource}

In this section, we evaluate the computational efficiency of our proposed \ours framework. Our focus is on the duration of each training epoch, particularly how it compares to scenarios without our specially designed module. To this end, we compare the training time of the private model in \ours with its counterpart that only uses symmetrical KD loss. Similarly, we assess the training time of the proxy model in \ours against training with only the CE loss. The average training time over three epochs is reported. The results in Table~\ref{tab:app_time} indicate that, while our approach does entail additional training time within a certain range, this increase is justifiable considering the performance improvements detailed in the ablation study (Table~\ref{tb:ablation}) and the communication cost reduction observed in Figure~\ref{fig:homo}. This aspect is also addressed in the discussion on limitations, where we contemplate strategies to further minimize computational costs in future work.